# Observation of vortex-pair dance and oscillation

Dadong Liu,[1] Lai Chen,[1] Li-Gang Wang [1,*]

[1] School of Physics, Zhejiang University, Hangzhou 310058, China
* Corresponding author. E-mail address: lgwang@zju.edu.cn (L.-G. Wang).

**Abstract**

Vortex dynamics, which encompass the motion, evolution, and propagation of vortices, elicit both fascination and challenges across various domains such as fluid dynamics, atmospheric science, and physics. This study focuses on fundamental dynamics of vortex-pair fields, specifically known as vortex-pair beams (VPBs) in optics. VPBs have gained increasing attention due to their unique properties, including vortex attraction and repulsion. Here, we explore the dynamics of pure-phase VPBs (PPVPBs) and observe intriguing helical and intertwined behaviors of vortices, resembling a vortex-pair dance. We uncover the oscillation property of the intervortex distance for PPVPBs in free space. The observed dancing and oscillation phenomena are intricately tied to the initial intervortex distance and can be explained well in the hydrodynamic picture. Notably, the vortex dancing and oscillation alter the process of vortex-pair annihilation, extending the survival range for opposite vortices. This discovery enhances our understanding of vortex interactions and sheds light on the intricate dynamics of both vortex-vortex and vortex-antivortex interactions.

**INTRODUCTION**

Vortices are prevalent phenomena observed across a spectrum of scales in nature, ranging from water eddies and atmospheric typhoon or hurricanes to majestic spiral galaxies. Vortices are also fundamental solutions within cylindrical-symmetry resonators for electromagnetic fields (*1*) and are pervasive in the realm of light, where they are known as optical vortices (*2, 3*). Optical vortices exhibit a distinctive feature— a dark core at the center, characterized by an indeterminate phase with vanishing amplitude (*4, 5*). These unique attributes of optical vortices give rise to a diverse array of applications, including optical micromanipulation (*6-8*), optical communications (*9-11*), quantum information (*12-15*), super-resolution imaging (*16-18*), and optical measurements (*19-21*).

In the presence of multiple vortices within light fields, the topological dynamics and interaction among vortices can create unique and interesting phenomena, like vortex knots (*22*), vortex collisions (*23, 24*), and the consequential process of vortex creation, annihilation or nucleation (*23-30*). Among vortex interactions, a fundamental scenario involves the interaction of two vortices. Like in atmosphere, Fujiwhara effect occurs as two typhoons approach each other (*31*). In the domain of optics, the dynamic interplay of attraction and repulsion between two vortices has been previously observed in the dynamics of vortex-pair fields (*25, 26*). A vortex-pair beam (VPB) is a type of structured light fields containing a pair of vortices. It is sometime categorized into two scenarios: an isopolar vortex pair, *i.e.* two vortices with identical topological charges (TCs), and a vortex dipole, characterized by two vortices with opposite TCs (*32*). In 1993, Indebetouw discovered that the relative distance of an isopolar vortex pair remains constant during free space propagation, while a vortex dipole tends to exhibit mutual attraction (*25*). This effect was subsequently confirmed through the experiment (*26*). The interaction between two vortices manifests specific features, including the rotational effect observed in the isopolar vortex pair during propagation (*26, 33*), the reappearance of an annihilated vortex dipole in the far field (*34*), and an optical intrinsic orbit–orbit interaction, serving as a manifestation of the attractive and repulsive interactions within a vortex dipole (*35*). Furthermore, the exploration of VPB dynamics in diverse optical systems (*36-49*), such as those involving a graded-index medium (*40, 41*), a high numerical-aperture lens (*42-44*), an astigatic system (*45-47*), a half-plane screen (*48*), and a knife edge (*49*), has been a subject of extensive discussion.

However, the aforementioned investigations (*25, 26, 32-49*) primarily rely on the model of complex-amplitude VPBs (CAVPBs), in which each vortex undergoes complex-amplitude modulation, constraining our comprehensive understanding of the dynamics inherent in multiple vortices. In the exploration of fractional vortex fields (*50-55*) and vortex arrays (*56-60*), researchers have found the intricate dynamics in the evolutions of vortex interactions among multiple vortices beyond the vortex attraction and repulsion process, like the birth or annihilation of vortex pairs. To further advance our understanding of the interaction between two vortices, here, we would like to address the dynamics of the pure-phase VPBs (PPVPBs), consisting of two pure-phase vortices. It is noteworthy that, to the best of our knowledge, the dynamics of PPVPBs have not been reported previously, despite their proposal and application in optical trapping and manipulation (*61*). Here, we elucidate the intriguing dynamics arising from vortex-vortex and vortex-antivortex interactions, resulting in a phenomenon reminiscent of vortex dance—a helical and intertwining behaviour among vortices. The observed vortex dynamics can be well explained by using an optical hydrodynamic picture (*62*). Our experimental verification, employing the interference method to trace vortex trajectories in light fields, solidifies the existence of this interesting feature. Notably, the oscillation of intervortex distance in PPVPBs represents a fundamentally unique characteristic not observed in traditional CAVPBs, where no oscillation effect had been previously identified. A comprehensive investigation reveals that the observed vortex dancing and oscillation can be precisely controlled by the initial intervortex distance, reflecting the interaction strength between the vortices. Meanwhile, the vortex dance and oscillation substantially influence the process of vortex-pair annihilation. This effect enlarges the survival range of opposite vortices, with a certain similarity to Fujiwhara effect of two typhoons in atmosphere, presenting a distinct aspect not witnessed in CAVPBs.

## RESULTS
### Fields of PPVPBs
We start by briefly reviewing the previous model of CAVPBs (*25*). The initial field of such CAVPBs embedded in a host Gaussian beam is usually expressed as

$$E_{\text{CAVPB}}(u,v) = M(u,v)\exp\left(-\frac{u^2+v^2}{w_0^2}\right) \quad \text{with}$$

$M(u,v) = [(u-u_0) + i\,\text{sgn}(m_1)v]^{|m_1|}[(u+u_0) + i\,\text{sgn}(m_2)v]^{|m_2|}$, where $2u_0$ is the initial distance of two vortices with TCs $m_1$ and $m_2$, $w_0$ is the beam width of the host Gaussian beam, and $u, v$ refer to the transverse rectangular coordinates at the initial plane. As the magnitude of the function $M(u, v)$ changes and deviates from unity, the two vortices in CAVPBs also undergo amplitude modulation. According to modal analysis (*32, 36*), these CAVPBs can be expressed as linear combinations of finite vortex modes, which, in turn, govern the dynamics of vortices. In most of works, researchers only considered the cases of $m_1 = m_2 = 1$ for an isopolar vortex pair or $m_1 = -m_2 = 1$ for a vortex dipole. The vortex trajectories, illustrating the attraction or repulsion of vortices, were demonstrated in a series of prior investigations (*25, 26, 32-49*). Recently, one has developed the laser hydrodynamic model to explain the vortex motions in multiple complex-amplitude vortices (*63, 64*). Nevertheless, unraveling the underlying physics of vortex dynamics remains a challenge in many complex optical fields, surpassing the complexity observed in CAVPBs.

In contrast to the model of CAVPBs, one can have an alternative choice on two pure-phase vortices, which can be written as (*61*)

$$E_{\text{PPVPB}}(u,v) = F(u,v)\exp\left(-\frac{u^2+v^2}{w_0^2}\right),$$

(1)

with $F(u,v) = \left[\dfrac{u-u_0+iv}{\sqrt{(u-u_0)^2+v^2}}\right]^{m_1}\left[\dfrac{u+u_0+iv}{\sqrt{(u+u_0)^2+v^2}}\right]^{m_2}$. Here the magnitude of $F(u, v)$ is always equal to unity, different from the above function $M(u, v)$, and Eq. 1 contains two pure-phase vortices like $e^{im_1\phi_1}$ and $e^{im_2\phi_2}$ with $\phi_1$ and $\phi_2$ being the two local azimuthal angles at $(\pm u_0, 0)$. Such fields are called as PPVPBs. In Fig. 1A, it shows the phase profiles of such PPVPBs for several situations with different $m_1$ and $m_2$. One can define the initial dimensionless relative off-axis distance $\tilde{u}_0 = u_0 / w_0$ as an indicator of external control on the interaction of vortex pair. It is noteworthy to reiterate that PPVPBs comprise two off-axis vortices with pure phase and without amplitude modulation. While this may appear similar to the aforementioned CAVPBs, it fundamentally differs from them. By examining the mode purities of PPVPBs and comparing them with CAVPBs, we ascertain that the orbital angular momentum (OAM) spectra of PPVPBs and CAVPBs are inherently distinct (refer to Section A of the Supplemental Materials). When $m_1 = m_2$, the OAM spectra of PPVPBs consist of infinite even OAM modes, whereas for CAVPBs their OAM spectra comprise finite even OAM modes. Conversely, when $m_1 = -m_2$, both PPVPBs and CAVPBs exhibit symmetrical OAM modes, with PPVPBs having an infinite set and CAVPBs having a finite set. These distinctions contribute to varied interactions between vortex-vortex and vortex-antivortex in PPVPBs, resulting in distinct behaviors of vortex dynamics.

The evolutions of optical fields in free space or an optical system can be well predicted by using theory of matrix optics (*65, 66*) and the detail description of theoretical equations can be found in the section of Materials and Methods. Once the field evolution is achieved, the locations of vortex centers can be determined either from the phase distributions or by identifying the dark cores through taking the logarithm of the light intensities, offering an intuitive display. The comparative movies between PPVPBs and CAVPBs are available in Section B of the Supplemental Materials. In the case of PPVPBs, their intensity distributions result in the formation of ripples in the light fields, akin to ripples on water caused by two falling stones, highlighting the interaction among vortices. In contrast, CAVPBs exhibit more stable and tranquil evolutions of intensity distributions during propagation, devoid of such ripples. We attribute these pronounced differences between PPVPBs and CAVPBs to the distinct dynamics of vortices, which could be seen from the vortex trajectories. We also show that the role of the host beam in PPVPBs is less important than that in CAVPBs, see the detail discussion in Section I of the Supplementary Materials.

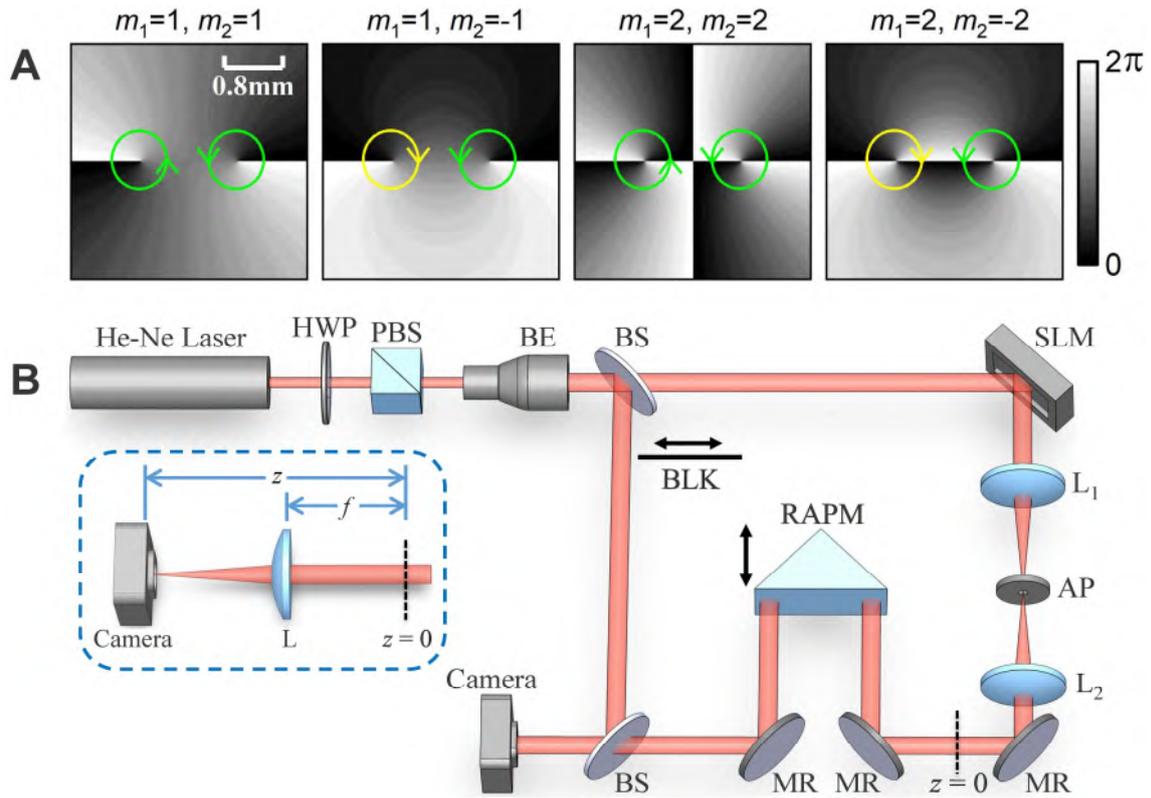

**Fig. 1. Schematic diagrams of phase distributions of PPVPBs and experimental setup.** (**A**) The initial phase distributions of PPVPBs with the TC values $m_1$ and $m_2$ marked on the top of subfigures. The phase singularities circled by green and yellow arrows represent positive and negative vortices, respectively. Other parameters here are $\tilde{u}_0 = 0.4$ and $w_0 = 1.63$ mm. (**B**) Experimental setup for generating PPVPBs and measuring the positions of phase singularities by the interference method. The position of $z=0$ is the generating plane of PPVPBs, which is also called as the input (or initial) plane for the subsequent optical system. Inset in (**B**) represents a specific focusing lens system with the lens' position located at $z=f$ from the input plane. Using this focusing system, at the back focal position of the lens (*i.e.*, $z=2f$ here), the system becomes a 2-*f* lens system and optical properties at $z=2f$ here is similar to the situation of the far-field or Fraunhofer region in free space. Notations are: HWP, half-wave plate; PBS, polarized beam splitter; BE, beam expander; BS, beam splitter; SLM, spatial light modulator; $L_1$ and $L_2$, the focusing lens with $f_1 = f_2 = 300$ mm; AP, aperture; MR, mirror reflector; BLK, block; RAPM, right-angle prism mirror. Here, the RAPM is movable for realizing the change of the propagation distance $z$ by using the electrically-controlled motorized system.

**Experimental setup**

To demonstrate the vortex dynamics, we experimentally generated PPVPBs by using a phase-only SLM (Holoeye PLUTO-2-NIR-015). Figure 1B depicts the schematic of our experimental setup designed to produce PPVPBs and detect vortex locations in free space using the interference method. We use a linearly-polarized He-Ne laser with a wavelength of 632.8 nm as the light source. The half-wave plate and the polarized beam splitter are used to control the horizontal polarization of the transmission light and adjust its light intensity. The beam was then expanded via a beam expander, increasing the beam waist ($w_0$) to approximately 1.63 mm. Subsequently, the beam underwent splitting by a beam splitter, with the reflected light serving as a reference beam for interference experiments, and the transmitted light is incident on the SLM to generate various orders of PPVPBs. Phase diagrams, as illustrated in Fig. 1A, were loaded onto the SLM. The modulated first-order diffraction beam, representing the generated PPVPB, was isolated using a suitable aperture. A 4-*f* lens system, composed of lenses $L_1$ and $L_2$ with focal lengths $f_1 = f_2 = 300$ mm, imaged the SLM plane onto the back focal plane of lens $L_2$. Consequently, PPVPBs were created on the rear focal plane of lens $L_2$, establishing the initial plane at $z = 0$ for studying the evolution of light fields in the subsequent optical system. A right-angle prism mirror, positioned on a motorized system, precisely adjusted the propagation distance ($z$) of the PPVPBs. Finally, the interference patterns between PPVPBs and the reference beam were captured by a camera with 12-bit depth. Experimentally obtained fringe patterns facilitated the reconstruction of PPVPB phase distributions using the Fourier-transform method (*67*). Based on this information, the vortex locations of the PPVPBs were determined through the application of the phase singularity search algorithm (*68*), leveraging the high-frequency characteristics of phase singularities. The methods to obtain the information of phase distributions and singularities are also introduced in the section of Materials and Methods.

**Dynamics of vortices in free space**
Now let us discuss the dynamics of vortex pair in the fields of PPVPBs. Figure 2 shows the trajectories of vortices for PPVPBs with $m_1 = m_2 = 1$ and $m_1 = -m_2 = 1$ in free space. When $m_1 = m_2 = 1$, the two positive vortices rotate individually, gradually repelling each other. Their trajectories exhibit central symmetry about the origin of the transverse plane, see their projection on the *x-y* plane. This centrosymmetric characteristic is independent of the propagation distance $z$ as shown in Fig. 2A and holds for all PPVPBs with equal TCs (*i.e.*, $m_1 = m_2$). In the case of $m_1 = -m_2 = 1$, both positive and negative vortices rotate themselves during propagation, but their trajectories exhibit symmetry about the *y*-axis. This reflectionally symmetric property remains unchanged across different propagation distance ($z$) as depicted in Fig. 2B. It is valid for all PPVPBs with opposite TCs (*i.e.*, $m_1 = -m_2$). These symmetries align with the symmetry properties of the initial phase distributions of such PPVPBs, which are symmetric about the origin or the *y*-axis, as displayed in Fig. 1A with $m_1 = m_2 = 1$ or Fig. 1A with $m_1 = -m_2 = 1$.

The evolution of each vortex within PPVPBs manifests more intriguing effects than those observed in CAVPBs. As depicted in Fig. 2 (A and B), the trajectory of each vortex in PPVPBs follows a helicoidal motion in free space. Simultaneously, the interplay among vortices induces oscillating changes in the intervortex distance, as evident in both experimental and theoretical results presented in Fig. 2 (C and D). This unique evolutionary pattern, involving simultaneous rotation and oscillation, resembles a dance and represents a interesting characteristic that has never previously observed in CAVPBs with linear polarization. Interestingly, the relative off-axis distance ($\tilde{u}_0$) plays a crucial role in the observed vortex oscillation phenomena. In Fig. 2 (C and D), it is noted that vortex oscillation persists for a longer propagation distance as $\tilde{u}_0$ increases. It is noteworthy that when $\tilde{u}_0$ is smaller than a certain value, vortices with opposite TCs can mutually annihilate each other after propagating a specific distance, see Fig. 2D. The critical value of $\tilde{u}_0$ for the annihilation feature in PPVPBs differs from that in CAVPBs. More information on the theoretical prediction on the intervortex distance can be found in Section D of the Supplementary Materials, and further instances of vortex annihilation phenomena in PPVPBs with opposite TC vortices also can refer to the videos in Section B of the Supplemental Materials. A quantitative comparison between PPVPBs and CAVPBs will be addressed in subsequent discussions.

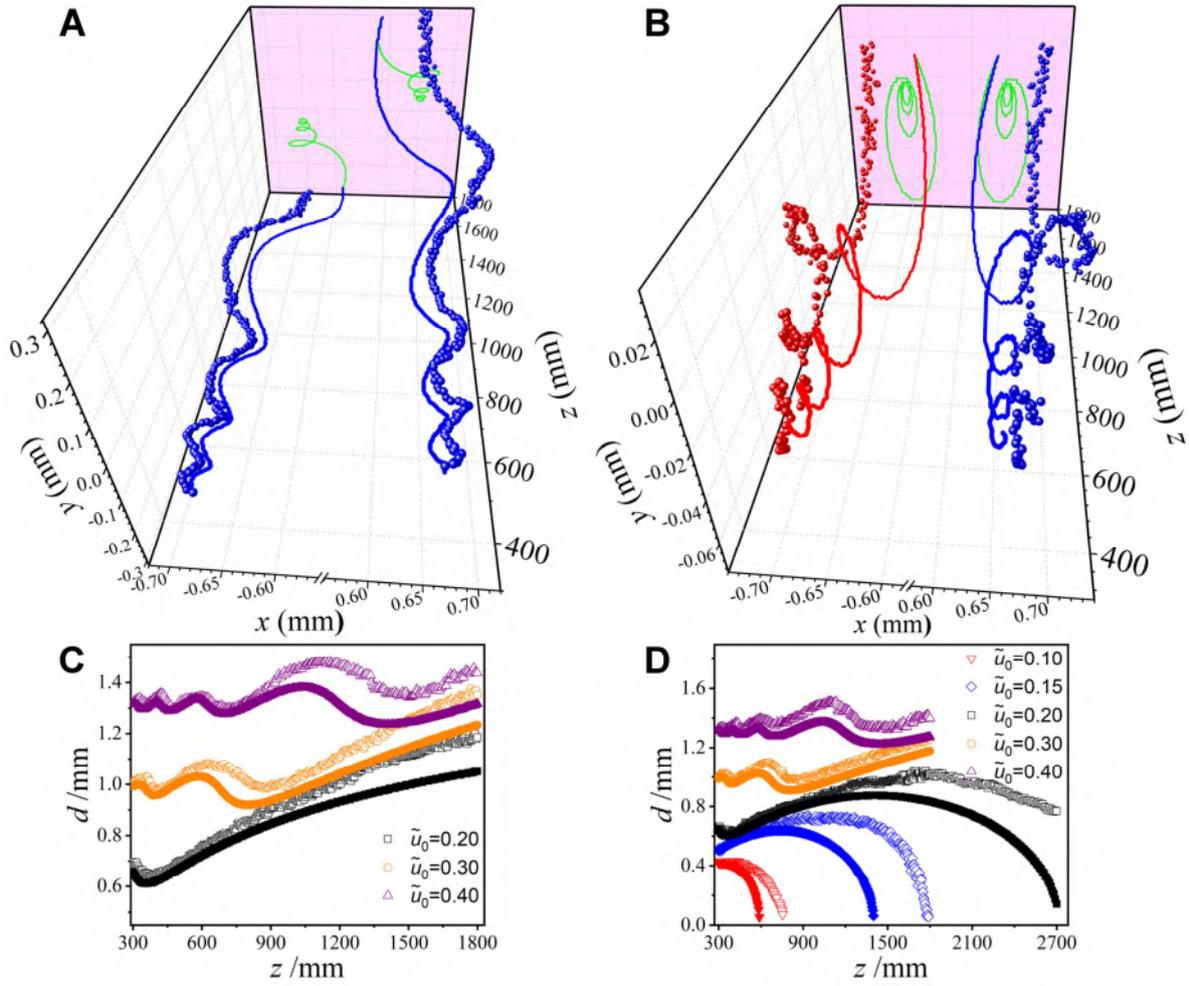

**Fig. 2. Experimental measurements of vortex trajectories and intervortex distance for PPVPBs in free space.** (**A** and **B**) Experimental vortex trajectories for (**A**) $m_1 = m_2 = 1$, and (**B**) $m_1 = -m_2 = 1$ with $\tilde{u}_0 = 0.4$. The blue and red dots denote, respectively, the evolution of positive and negative vortices. The corresponding solid lines are theoretical predictions and their projections are shown by the green lines in the $x$-$y$ planes. (**C** and **D**) Evolution of the intervortex distance along the propagation distance for (**C**) $m_1 = m_2 = 1$ and (**D**) $m_1 = -m_2 = 1$, respectively, under different $\tilde{u}_0$. The corresponding theoretical predications are also displayed with the same-colour curves. The experimental parameter $w_0 = 1.63$ mm is taken for theoretical calculations.

Figure 3 further presents both experimental and theoretical trajectories of vortices in the fields of PPVPBs with equal TCs ($m_1 = m_2 = 2$) and opposite TCs ($m_1 = -m_2 = 2$) in free space. As shown in Fig. 1A, the PPVPB with $m_1 = m_2 = 2$ showcases a central symmetry of two vortices bearing +2 TC each, which progressively split into four distinct vortices with individual TCs of +1 during propagation. From Fig. 3A, an interesting interplay among vortices emerges, entwining them in a helicoidal dance as the propagation distance $z$ extends. Intriguingly, although no oppositely signed vortices are present in the initial light field with $m_1 = m_2 = 2$, the evolving $z$ engenders the generation of multiple pairs of positive and negative vortices, engaging in a mesmerizing alternation of intertwining, nucleation, and annihilation.

For PPVPBs with opposite TCs ($m_1 = -m_2 = 2$), as depicted in Fig. 3B, even when a pair of vortices with ±2 TCs is initially present on the plane (refer to Fig. 1A), these vortices gracefully split into two pairs, one carrying +1 TCs and the other -1 TCs. Evidently, vortices boasting high-order TCs in PPVPBs exhibit instability, consistently fragmenting into vortices with +1 or -1 TC. Analogy to the phenomena observed in PPVPBs with equal TCs ($m_1 = m_2 = 2$), the entanglement, helical dynamics, and the intriguing nucleation and annihilation of vortices are also observed in PPVPBs with opposite TCs ($m_1 = -m_2 = 2$).

Based on the theory of optical hydrodynamics in the recent studies (*62-64*), the motion of vortices, the splitting of higher-charge optical vortices, and the dynamics of vortices have been explained in complex-amplitude modulation vortex pairs, in which no helicoidal, intertwined and oscillating vortex dynamics have been observed. Here, we employ the same argument to explain the vortex dynamics, especially in PPVPBs. The initial total light field consisting of vortex pairs embedded in a host Gaussian beam can be written as a product of two fields: one for the initial tested field of one vortex under consideration, another for the initial background for the rest field that comprises the fields of other vortices and the host beam. The evolutions of the transverse velocity fields of the background field at any propagation distance $z$ are demonstrated in the supplementary movies S3 and S4. In the movies, the tested vortex moves along the direction of the velocity field. In the PPVPBs with $m_1=m_2=1$ or $m_1=-m_2=1$, the tested vortex "surfs" in the diffraction waves from the other vortices in the background fields that induces the helical and oscillating motions (that explains the trajectories in Fig. 2). In the PPVPBs with higher-order TCs, the presence of the circulation flow from the vortex near the tested one further alters the local background velocity field. Under the diffraction ripples of the background field during propagation, the tested vortices experience not only the helical and oscillating motions but also the vortex nucleation and annihilation phenomena. In contrast, there are no complex velocity flows in CAVPBs (lacking the diffraction waves from other vortices), so there are no helical and oscillating motions of vortices. More information on the theoretical consideration of the optical hydrodynamic model and detail discussion on the supplementary movies S3 and S4 can be found in Section C of the Supplementary Materials.

From the above, the demonstrated helical and intertwined behaviors among vortex pairs not only induce the oscillation of the intervortex distance but also change the dynamics of vortex interactions like prolong the survival range of opposite vortices as discussed later. The characteristics of the dynamic behaviors within these PPVPBs intensifies with the augmentation of the relative off-axis distance ($\tilde{u}_0$). We can image that when $\tilde{u}_0$ is large, the tested vortex is immersed for more "time" in the diffracted waves of other vortices and oscillates in a spiral form. Sections E and F of the Supplementary Materials provide further results, confirming the influential role of $\tilde{u}_0$ in regulating vortex behaviors. Consequently, the relative off-axis distance ($\tilde{u}_0$) emerges as a pivotal control parameter for modulating the dynamic behaviors exhibited by these PPVPBs.

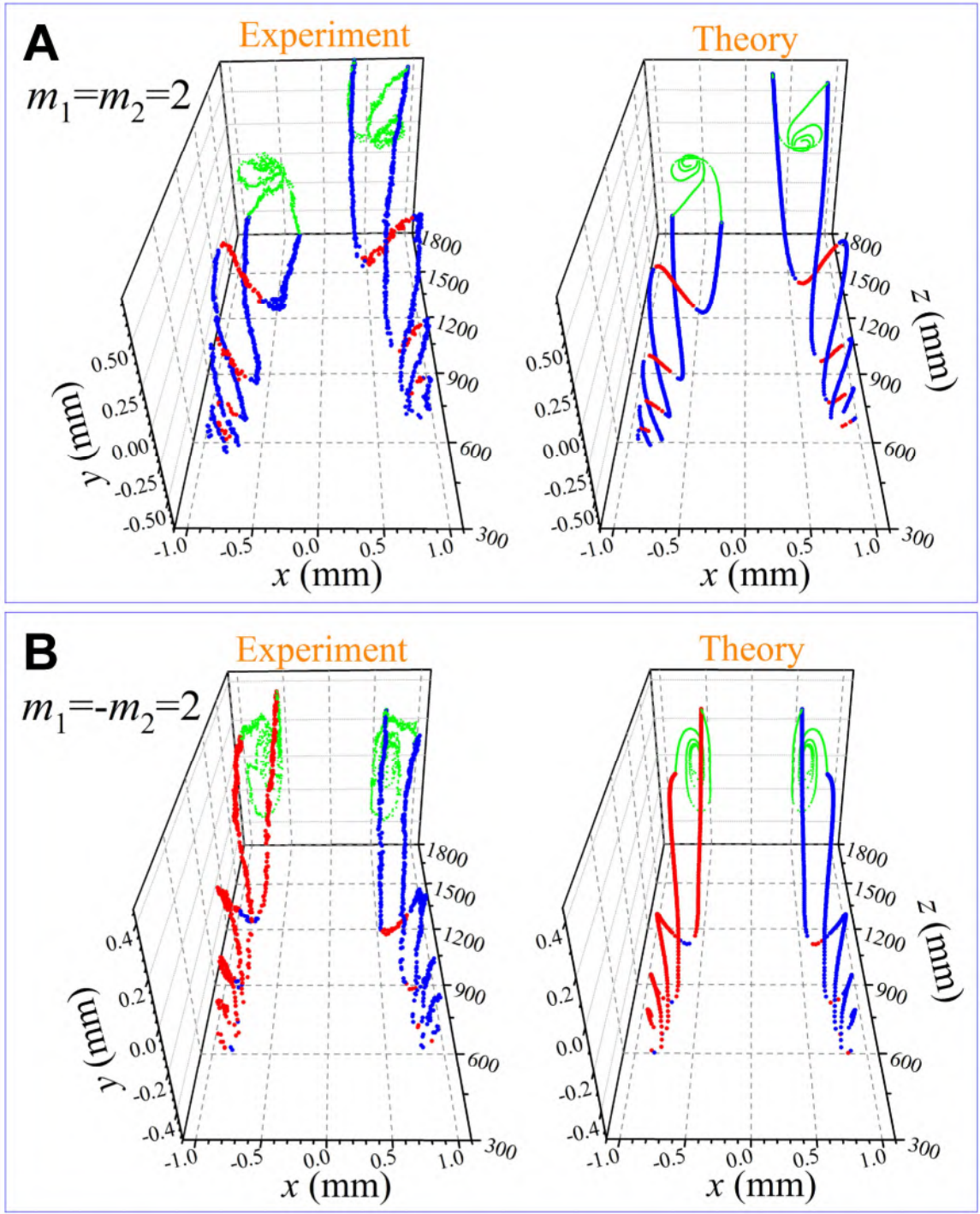

**Fig. 3. Experimental and theoretical trajectories of vortices in PPVPBs propagating in free space.** (**A**) The vortex trajectories for the PPVPB with equal TCs of $m_1 = m_2 = 2$ and (**B**) the vortex trajectories for the PPVPB with opposite TCs of $m_1 = -m_2 = 2$. Here the relative off-axis distance parameter is taken as $\tilde{u}_0 = 0.4$. The blue and red dots denote, respectively, data for the trajectories of positive and negative vortices, and their projections in the *x-y* plane are presented by the green dots. The experimental parameter $w_0 = 1.63$ mm is also taken for theoretical calculations.

## Dynamics of vortices in a focusing system

It is widely recognized that the light field in far-field regions closely resembles that found at the back focal plane of a 2-*f* lens system (*50*). Therefore, to thoroughly explore the annihilation process of vortices for the PPVPBs in the far field, it is more convenient to examine the evolution within the 2-*f* lens system illustrated in the inset of Fig. 1B. Through changing the value of *z* in this focusing system, one can achieve the key dynamics of vortex pairs from the near-field region (*i.e.*, the lens plane) to the far-field region (*i.e.*, the focal plane). The only difference is that the beam profile of light in free space is spread out or divergent while it becomes to be condensed or convergent in the focusing system.

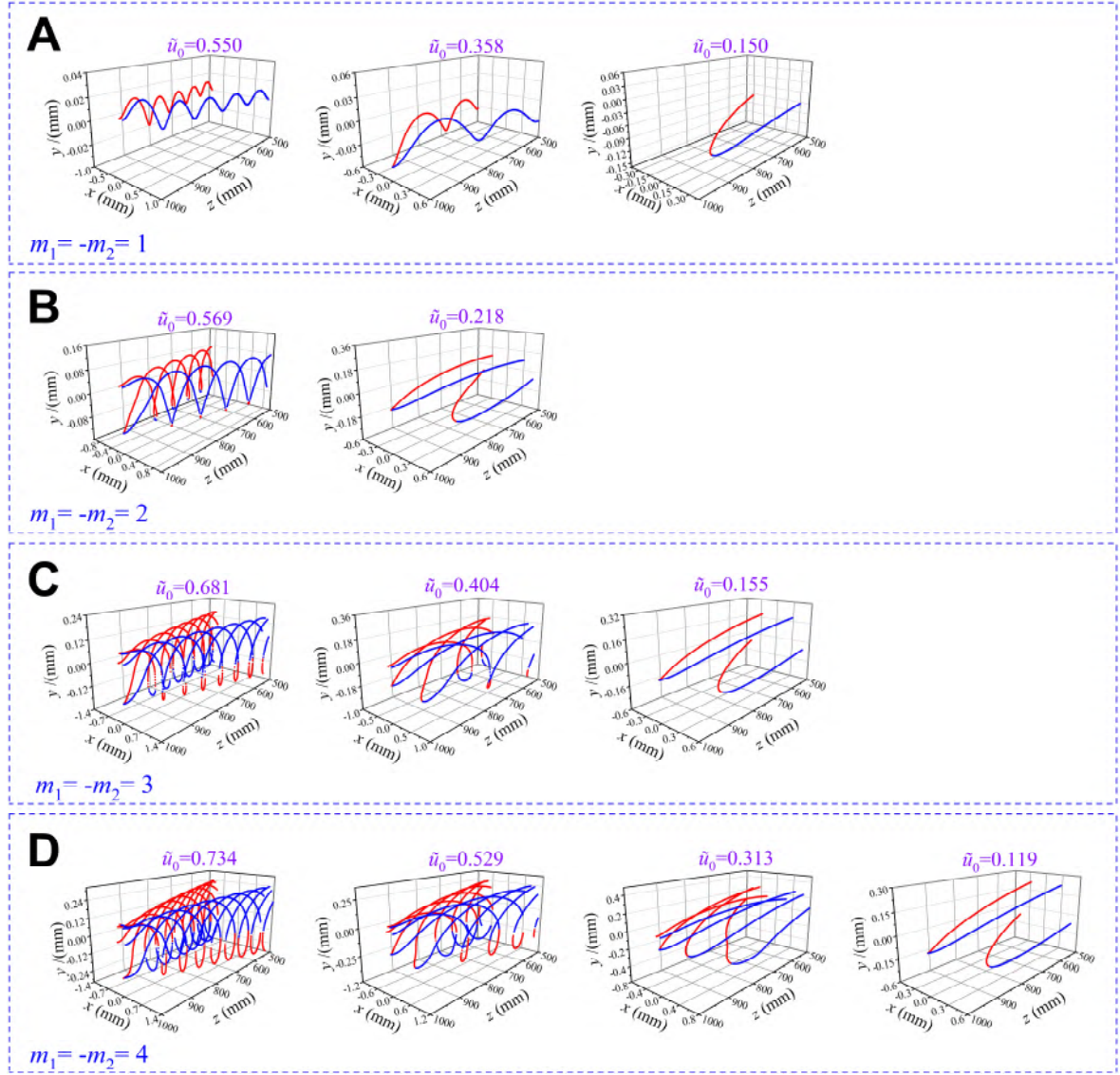

**Fig. 4. Evolution of vortex trajectories for different-order PPVPBs with opposite TCs in the 2-*f* focusing system.** (**A**) The influence of the relative off-axis distance $\tilde{u}_0$ on the vortex-trajectory evolution, where the opposite vortices happen to merge and annihilate each other at the focusing plane when $\tilde{u}_0 = 0.358$. (**B**, **C**, and **D**) The vortex-trajectory evolutions at various critical values of $\tilde{u}_0$ for different-order PPVPBs, in which each plot corresponds to the situation that one pair of opposite vortices annihilate each other at the focusing plane. Here, the focal length of the 2-*f* focusing system is *f* = 500 mm and the beam parameter is also taken to be $w_0 = 1.63$ mm.

In Fig. 4A, the influence of the relative off-axis distance $\tilde{u}_0$ on the trajectories of vortices, when $m_1 = -m_2 = 1$, is depicted as they evolve from the lens (*z*=500 mm) to the back focal plane (*z*=2*f*=1000 mm). When the value of $\tilde{u}_0$ is sufficiently large, such as when $\tilde{u}_0 = 0.55$, the positive and negative vortices undergo oscillations and persist at the back focal plane. This observation suggests that they do not annihilate each other in free space. Conversely, when $\tilde{u}_0 = 0.358$, the positive and negative vortex pair coincidentally merge at the focal plane, signifying their annihilation in the infinity of free space. This particular value is termed the critical value for the occurrence of vortex annihilation in PPVPBs when $m_1 = -m_2 = 1$. As $\tilde{u}_0$ decreases further, the annihilation phenomenon happens before the focal plane, indicating that this effect can be observed at a suitable distance in free space. In the Section H of the Supplementary Materials, we also provide the corresponding change of the intervortex distance of the vortex pair in free space. Notably, the critical value of $\tilde{u}_0$ in PPVPBs is much smaller than that in cases of CAVPBs with $m_1 = -m_2 = 1$. This distinction implies that the dynamic properties of opposite vortices in PPVPBs during evolution surpass those of CAVPBs under identical conditions. From a propagation perspective, owing to the oscillatory or dancing behaviors between opposite vortices in PPVPBs, the annihilation process becomes much slower, allowing vortices to survive over longer distances. Consequently, a slight increase in $\tilde{u}_0$ results in the disappearance of the annihilation process compared to that in CAVPBs. This effect in PPVPBs bears similarity to the Fujiwhara effect between two typhoons, which often prolongs the lifespan of typhoons (*69*).

For $m_1 = -m_2 = 2$, the initial light field processes a pair of opposite vortices with ±2 TCs, but due to the unstable properties of high-order vortex pair during propagation, they rapidly split into two pairs of opposite vortices with ±1 TCs, thus there are complex dynamics as shown in Fig. 3B. In Fig. 4B, two critical situations for vortex annihilation at the focal plane are shown. When $\tilde{u}_0 = 0.569$, one pair undergoes annihilation at the focal plane, while the other pair survives. When $\tilde{u}_0$ further reduces to $\tilde{u}_0 = 0.218$, the second pair also undergoes annihilation at the focal plane. Interestingly, at this situation the first pair of vortices actually annihilate each other at a shorter distance or earlier. Note that the phenomenon of vortex dancing here appears before the lens plane. Thus, for cases when $m_1 = -m_2 = 2$, there are two critical values of $\tilde{u}_0$, each corresponding to the critical points of annihilation processes for the respective pairs of vortices.

Similarly, as demonstrated in Fig. 4(C and D), for the instances of $m_1 = -m_2 = 3$ and $m_1 = -m_2 = 4$, three and four pairs of opposing vortices with ±1 TCs are observed, respectively. Each scenario is associated with a distinct critical value corresponding to the annihilation of a vortex pair at the focal plane. Additionally, within Fig. 4, the focal fields of PPVPBs reveal the oscillation and dancing of vortex trajectories, featuring vortex intertwining and helical behaviors, phenomena hitherto unobserved in CAVPBs.

**Table 1. Comparison of critical values of the relative off-axis distance $\tilde{u}_0$ between PPVPBs and CAVPBs under different-order opposite TCs for occurring vortex-pair annihilation at the focal plane of a 2-*f* lens system.**

| | ±1 TCs | ±2 TCs | | ±3 TCs | | | ±4 TCs | | | |
|---|---|---|---|---|---|---|---|---|---|---|
| PPVPBs | 0.358 | 0.569 | 0.218 | 0.681 | 0.404 | 0.155 | 0.734 | 0.529 | 0.313 | 0.119 |
| CAVPBs | 0.500 | 0.925 | 0.383 | 1.254 | 0.758 | 0.323 | 1.533 | 1.066 | 0.661 | 0.285 |

Table 1 enumerates critical values of $\tilde{u}_0$ for observing the annihilation effect of each pair of opposite vortices with ±1 TCs within PPVPBs at the focal plane in the cases of $m_1 = -m_2$. For comparison, corresponding critical values for observing annihilation effects in CAVPBs at the focal plane are also included. Interestingly, each critical value of $\tilde{u}_0$ for PPVPBs is smaller than the corresponding critical value for CAVPBs. In other words, under identical conditions (for example, with the same value of $\tilde{u}_0$ and beam parameters), the annihilation effect occurs at a greater distance for PPVPBs than for CAVPBs. This delay is attributed to the vortex oscillation and dancing effects, which prolong the annihilation process of vortex pairs. Additional details regarding the annihilation processes of vortices in CAVPBs with opposite TCs in a 2-*f* lens system are provided in Section G of the Supplemental Materials.

**DISCUSSION**

Our study unveils the dynamic behaviors of PPVPBs in both free space and the focusing system. For PPVPBs featuring unit vortices, vortex trajectories form helical structures, accompanied by oscillating intervortex distances between vortices. Our experimental results are well demonstrated and confirm the theoretical predictions. In the case of PPVPBs with high-order vortices, the initial high-order vortices undergo a dynamic process, splitting into multiple unit vortices during propagation. This evolution is marked by intricate vortex intertwining and helical behaviors including vortex nucleation and annihilation. Such fast helical and intertwining behaviors resemble a dance of vortex pairs and are very common in the fields of PPVPBs. Interestingly, the light fields of PPVPBs with high-order TCs exhibit the nucleation, evolution, and annihilation of positive and negative vortex pairs during propagation. The vortex intertwined, and helical dance of vortices evoke a visual effect, resembling multiple pairs of vortex dancing. The observed vortex dynamics are explained physically from the hydrodynamics of light fluids. Notably, the intervortex distance between the vortex pair at the initial plane serves as a control parameter for orchestrating vortex dancing. This vortex dancing, in turn, emerges as the primary interaction driving the prolongation of the annihilation process for vortices with opposite TCs. These results underscore the distinctiveness of vortex dynamics in PPVPBs compared to CAVPBs. Our findings offer deeper insights into vortex interactions and hold potential applications in optical micromanipulation and the transportation of optical vortex information.

## MATERIALS AND METHODS

### Evolutions of optical fields in paraxial systems

Here, we employ theory of light diffraction in the paraxial approximation. The evolution of PPVPBs through a linear *ABCD* optical system, such as free space or a lens system, can be theoretically predicted from the Collins formula (*65, 66*)

$$E_o(x,y,z) = \frac{\exp(ikz)}{i\lambda B} \iint E_i(u,v) \exp\left\{\frac{ik}{2B}[A(u^2+v^2) - 2(xu+yv) + D(x^2+y^2)]\right\} du dv \quad (2)$$

where *A*, *B*, and *D* denote the elements of the ray transfer matrix $\begin{pmatrix} A & B \\ C & D \end{pmatrix}$ for a linear optical system, $k = 2\pi/\lambda$ is the wave number, $\lambda$ is the wavelength, and *z* is the propagation distance along the propagation axis. The light field of PPVPBs at the output plane (*i.e.*, the observation plane) is obtained by substituting Eq. 1 into Eq. 2. In free space propagation, the ray transfer matrix is represented as (*52*) $\begin{pmatrix} A & B \\ C & D \end{pmatrix} = \begin{pmatrix} 1 & z \\ 0 & 1 \end{pmatrix}$, where *z* denotes the propagation distance in free space. For the beam propagation in a 2-*f* lens system, the ray transfer matrix is expressed as (*66*) $\begin{pmatrix} A & B \\ C & D \end{pmatrix} = \begin{pmatrix} 1-(z-f)/f & f \\ -1/f & 0 \end{pmatrix}$, with *f* denoting the focal length of the 2-*f* lens system, *z* as the propagation distance from the input to the output plane, and the lens located at *z = f*. In the 2-*f* lens system, it is well known that Eq. 2 becomes the two-dimensional (2D) Fourier transformation, since both *A* and *D* are equal to zero and *B=f* at the back focal position of the 2-*f* lens system. This 2D Fourier transformation is similar to the Fraunhofer diffraction equation of light at the far-field region of free space (*50*). The purpose of using the 2-*f* lens system here is to conveniently investigate the far-field behavior. To visually represent the intensity evolution of the light fields, Eq. 2 provides a theoretical basis.

### Methods of achieving the phase distributions and phase singularities

Accurate positioning of vortices in PPVPBs requires the phase information of optical fields. According to the Collins formula (*i.e.*, Eq. 2) and the ray transfer matrix, one can theoretically obtain the phase distributions of PPVPBs in free space or the 2-*f* lens system. On that basis, one can attain the theoretical locations of vortices for the PPVPBs during propagation by using the vortex location algorithm in Ref. (*68*), which is based on the high-frequency characteristics of vortices. Thus, theoretical vortex positions can be achieved by using the matrix-optics theory and the vortex search approach. In experiments, accurate measurement of vortices is more complex, since we can only directly record the intensity distribution of the light field, rather than the phase distribution. Although it is possible to determine the vortex location through the dark region of light intensity, the accuracy of this method is relatively low compared to the vortex search algorithm based on phase information (*68*). In order to locate accurately the position of vortices in the experiment, we should firstly attain the experimental phase information of the light field. To reconstruct the experimental phase distribution of PPVPBs, we can use phase recovery methods (*67*). The principle of the phase recovery method in Ref. (*67*) is mainly based on the Fourier-transform of interference patterns. On the basis of the reconstructed phase distributions and the vortex search algorithm in Ref. (*68*), we can obtain the experimental vortex locations of PPVPBs.

## Supplementary Materials
**This PDF file includes:**
Supplementary Text
Figs. S1 to S18
Legends for movies S1 to S4

**Other Supplementary Material for this manuscript includes the following:**
Movies S1, S2, S3 and S4 (please download from: https://github.com/wangligangZJU/video )

**Acknowledgments:** L.-G. W. thanks Prof. C. T. Chan and Prof. Zhaoqing Zhang at Hong Kong University of Science and Technology for valuable discussions.

**Funding:** This work was supported by the National Natural Science Foundation of China (nos. 62375241 and 11974309).

**Author contributions:** Conceptualization: L.G.W. Investigation: D.L., L.C., L.G.W. Formal analysis: D.L, L.C., L.G.W. Visualization: D.L., L.C. Validation: D.L., L.G.W. Writing-original draft: D.L. Writing-review & editing: D.L., L.C., L.G.W.

**Competing interests:** The authors declare that they have no competing interests.

**Data and materials availability:** All data needed to evaluate the conclusions in the paper are present in the paper and/or the Supplementary Materials.

# Supplementary Materials for
# **Observation of vortex-pair dance and oscillation**

Dadong Liu *et al.*

Corresponding author: Li-Gang Wang, lgwang@zju.edu.cn

**This PDF file includes:**
    Supplementary Text
    Figs. S1 to S18
    Legends for movies S1 to S4

**Other Supplementary Material for this manuscript includes the following:**
    Movies S1, S2, S3 and S4 (please download from: https://github.com/wangligangZJU/video )

## A. Orbital angular momentum (OAM) spectra of pure-phase vortex-pair beams (PPVPBs) and complex-amplitude vortex-pair beams (CAVPBs) at the initial plane

The helical harmonic exp($il\varphi$) is the eigenfunction of orbital angular momentum (OAM), where $l$ is the topological charge (TC), and $\varphi$ denotes the azimuthal coordinate (70). A light field $E(r,\varphi,z)$ can be expanded through helical harmonics exp($il\varphi$) as (71, 72)

$$E(r,\varphi,z) = \frac{1}{\sqrt{2\pi}} \sum_{l=-\infty}^{+\infty} a_l(r,z) \exp(il\varphi), \quad (S1)$$

where the expansion coefficients $a_l(r,z)$ are given by

$$a_l(r,z) = \frac{1}{\sqrt{2\pi}} \int_0^{2\pi} E(r,\varphi,z) \exp(-il\varphi) d\varphi. \quad (S2)$$

The intensity of the $l$-th order helical harmonic, which is usually independent of the parameter $z$, can be expressed as

$$C_l = \int_0^{+\infty} |a_l(r,z)|^2 r dr. \quad (S3)$$

Thus, the intensity weight of such helical harmonic is determined by

$$R_l = \frac{C_l}{\sum_{q=-\infty}^{+\infty} C_q}, \quad (S4)$$

which can be seen as the OAM spectra or mode purities of the light field $E(r,\varphi,z)$ (73, 74).

The light field of the PPVPBs at the initial plane is given by

$$E_{PPVPB}(u,v) = F(u,v) \exp\left(-\frac{u^2+v^2}{w_0^2}\right), \quad (S5)$$

with $F(u,v) = \left[\frac{u-u_0+iv}{\sqrt{(u-u_0)^2+v^2}}\right]^{m_1} \left[\frac{u+u_0+iv}{\sqrt{(u+u_0)^2+v^2}}\right]^{m_2}$, where $m_1$ and $m_2$ are the TC values for vortices. The two vortices are located at $(u_0,0)$ and $(-u_0,0)$, respectively. In general, these two vortices can be, respectively, located at $(u_0,v_0)$ and $(-u_0,-v_0)$. Due to the rotational symmetry, one can make any angle of rotation to align these two vortices along the $u$ axis. Therefore, for the sake of simplicity, we only consider the situation of two vortices aligned along the $u$ axis without loss of generality.

The light field of the CAVPBs at the initial plane is expressed as

$$E_{CAVPB}(u,v) = M(u,v) \exp\left(-\frac{u^2+v^2}{w_0^2}\right), \quad (S6)$$

with $M(u,v) = [(u-u_0)+i\text{sgn}(m_1)v]^{|m_1|}[(u+u_0)+i\text{sgn}(m_2)v]^{|m_2|}$, for CAVPBs with equal or opposite TCs respectively (25).

Clearly, the amplitude of $M(u, v)$ in CAVPBs is not normalized, while the amplitude of $F(u, v)$ in PPVPBs is normalized. From Eq. (S6), the two vortices in the CAVPBs undergo not only the phase modulation but also the amplitude modulation. There are a series of prior investigations (25, 26, 32-49) on such fields of CAVPBs, demonstrating the attraction or repulsion of two vortices. However, it is hard to know the role of pure phase-only modulation on the dynamics of vortices. In Eq. (S5), for the fields of PPVPBs, it can be rewritten as $F(u,v) = e^{im_1\phi_1} e^{im_2\phi_2}$ with the two local azimuthal angles $\phi_1 = \arctan(\frac{v}{u-u_0})$ and $\phi_2 = \arctan(\frac{v}{u+u_0})$ centered at $(\pm u_0, 0)$. These two vortices suffer purely phase modulations. Under this model, one can explore the role of two pure phase-modulated vortices on their dynamic behavior. One may think that Eqs. (S5) and (S6) look to be similar each other, but this study shows that the different forms of these two kinds of vortex pairs result in distinct behaviors in their dynamics. For example, the helical and intertwined behaviors exhibited in the dynamics of PPVPBs, resembling a specific dance of vortex pairs, have *never* noticed in the previous studies on CAVPBs. Meanwhile, the impact of the host Gaussian beam on PPVPBs and CAVPBs is provided in Section I.

It is not difficult to rewrite, respectively, both Eq. (1) (or Eq. (S5)) and Eq. (S6) into the cylindrical coordinate system as

$$E_{PPVPB}(r,\varphi) = F(r,\varphi)\exp\left(-\frac{r^2}{w_0^2}\right)$$
$$= \exp\left(-\frac{r^2}{w_0^2}\right)e^{im_1\phi_1}e^{im_2\phi_2},$$
(S7)

and

$$E_{CAVPB}(r,\varphi) = M(r,\varphi)\exp\left(-\frac{r^2}{w_0^2}\right)$$
$$= (r^2+u_0^2-2u_0r\cos\varphi)^{\frac{m_1}{2}}(r^2+u_0^2+2u_0r\cos\varphi)^{\frac{m_2}{2}}\exp\left(-\frac{r^2}{w_0^2}\right)e^{im_1\phi_1}e^{im_2\phi_2},$$
(S8)

where $\phi_1 = \arctan(\frac{v}{u-u_0}) = \arctan(\frac{r\sin\varphi}{r\cos\varphi-u_0})$ and $\phi_2 = \arctan(\frac{v}{u+u_0}) = \arctan(\frac{r\sin\varphi}{r\cos\varphi+u_0})$ are the two local azimuthal angles centered at $(u_0,0)$ and $(-u_0,0)$, respectively. We can drop the variable $z$ in Eqs. (S1)-(S3) since we only consider the initial fields. According to Eq. (S2), from the mathematical point of view, the expansion coefficient $a_l(r)$ for the helical harmonic $\exp(il\varphi)$ is strongly dependent on their initial field expressions $E_{PPVPB}(r,\varphi)$ and $E_{CAVPB}(r,\varphi)$. We have confirmed via the numerical calculations that the expansion coefficient $a_l(r)$ is real for the initial fields of both Eqs. (S7) and (S8).

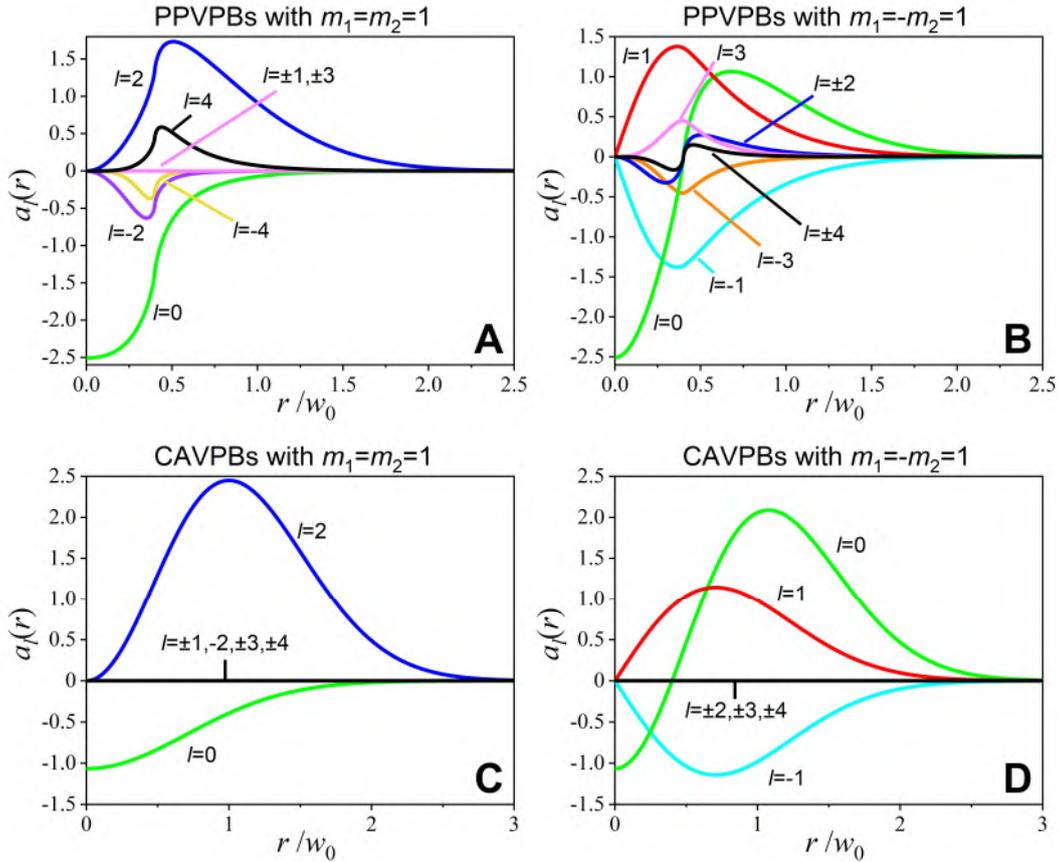

**Fig. S1. The expansion coefficients $a_l(r)$ of helical harmonics $\exp(il\varphi)$ within $-4 \leq l \leq 4$ for both PPVPBs and CAVPBs.** (A and B) The PPVPBs with $m_1 = m_2 = 1$ and $m_1 = -m_2 = 1$ and (C and D) the CAVPBs with $m_1 = m_2 = 1$ and $m_1 = -m_2 = 1$. Note that in (A), the odd helical harmonics disappear since $a_l(r)=0$ for $l = \pm 1, \pm 3$; in (B), the even helical harmonics overlap each other for $l = \pm 2$ and $l = \pm 4$, respectively; in (C), only two helical harmonics exist and other modes are zero; in (D), only three helical modes appear and other modes are absent. Here we emphasize that there are other higher order modes for the PPVPBs since the range of modal orders is limited within $-4 \leq l \leq 4$ for illustration. The parameters are $\tilde{u}_0 = 0.4$ and $w_0 = 1.63$ mm.

In Fig. S1, we plot the distributions of different $a_l(r)$ within $-4 \leq l \leq 4$ for both PPVPBs and CAVPBs in the cases of $m_1 = m_2 = 1$ and $m_1 = -m_2 = 1$. It demonstrates clearly that for the field of the PPVPB with $m_1 = m_2 = 1$, see Fig. S1A, the odd modes have no contribution while the even modes have significant contributions and constitute the field of the PPVPB; for the CAVPB with $m_1 = m_2 = 1$ as shown in Fig. S1C, it is only composed of two modes with $l = 0, 2$ and all other modes are not present. In the case of $m_1 = -m_2 = 1$, all the modes have contributions for the PPVPB while there are only the contributions from the three modes with $l = 0, \pm 1$ for the CAVPB. Obviously, the additional amplitude modulation in $E_{\text{CAVPB}}(r,\varphi)$ not only changes the distributions of $a_l(r)$ but also strongly suppresses the contributions of other helical harmonics.

Figure S2 shows OAM spectra of the PPVPBs and CAVPBs with equal TCs $m_1 = m_2 = +1$ under different relative off-axis distances $\tilde{u}_0$ at the initial plane. As the value of $\tilde{u}_0$ increases, the intensity weights of the zero-order OAM spectra for both PPVPBs and CAVPBs increase. However, for equal TCs +1, the OAM spectra of PPVPBs contains infinite even OAM modes, while CAVPBs consists of finite even OAM modes (i.e. the zero- and second-order OAM modes). To better understand the properties of OAM spectra of vortex pairs, we plot Fig. S3 to demonstrate the distributions of OAM spectra for PPVPBs and CAVPBs with different TCs. For high-order equal TCs, such as $m_1 = m_2$ = +2 and $m_1 = m_2$ = +3, the properties of the OAM spectra for PPVPBs and CAVPBs composed of infinite and finite even OAM modes, respectively, are similar to those of PPVPBs and CAVPBs with equal TCs +1. Interestingly, for opposite TCs, the distributions of OAM spectra for both PPVPBs and CAVPBs are symmetric about TC $l = 0$, but infinite OAM modes for PPVPBs and finite OAM modes for CAVPBs.

Through the detail analysis of theoretical calculations, we find that the fields of CAVPBs consist of finite helical harmonic modes with $l = 0, 2, \cdots, 2m$ in the cases of $m_1 = m_2 = m$, or $l = 0, \pm 1, \pm 2, \cdots, \pm m$ in the cases of $m_1 = -m_2 = m$ (here $m > 0$), while the fields of PPVPBs always comprise a series of infinite helical harmonic modes with $l = 0, \pm 2, \pm 4, \cdots$ in the cases of $m_1 = m_2 = m$, or $l = 0, \pm 1, \pm 2, \pm 3, \cdots$ in the cases of $m_1 = -m_2 = m$.

Since the only difference between PPVPBs and CAVPBs is whether there is amplitude modulation or not, we argue that the mode difference between PPVPBs and CAVPBs is probably the main reason for inducing the different dynamics of PPVPBs from that of CAVPBs. However, there are still some mysterious and deep questions that remain unresolved, such as how amplitude modulation affects the OAM spectra of PPVPBs and CAVPBs, and how the OAM spectra correlate with vortex dynamics, which should be further investigated in the future.

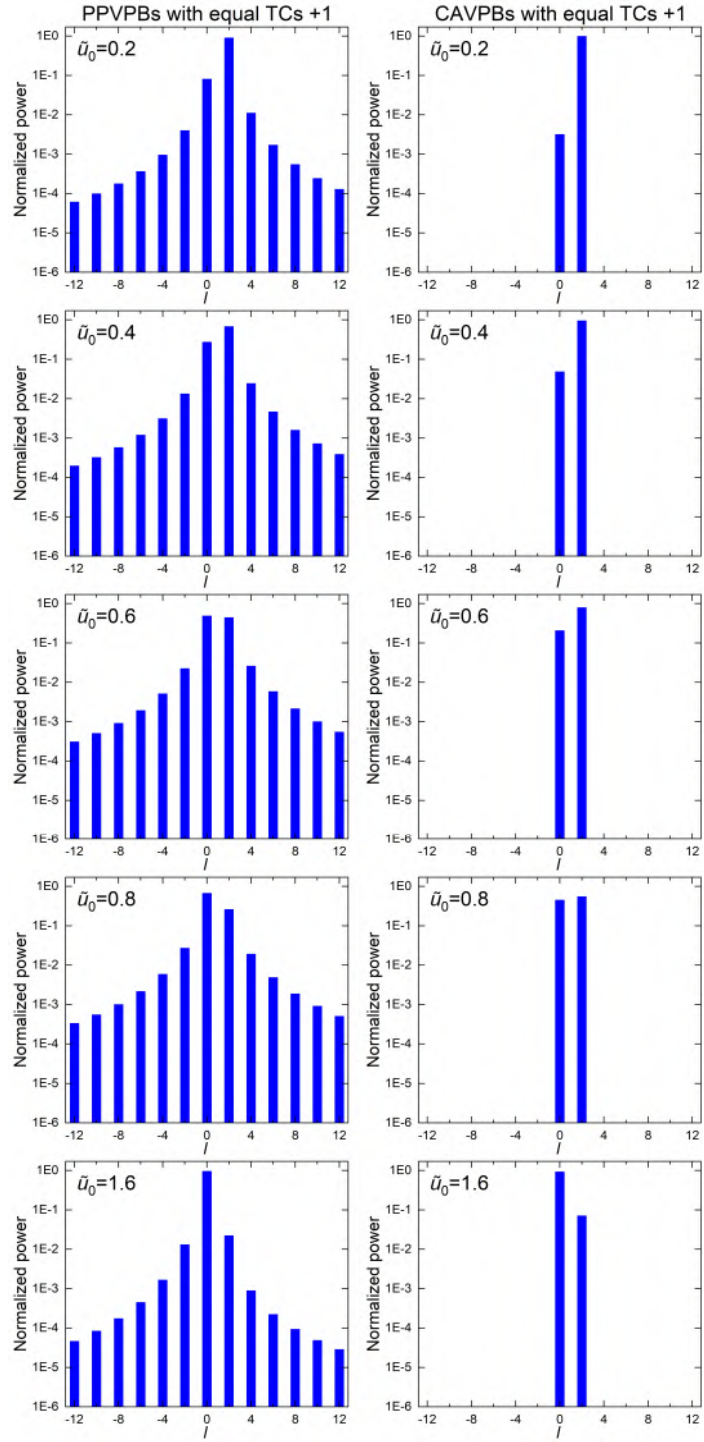

**Fig. S2. OAM spectra for the PPVPBs and CAVPBs with equal TCs +1 under different $\tilde{u}_0$ at the initial plane.** The parameter is $w_0 = 1.63$ mm.

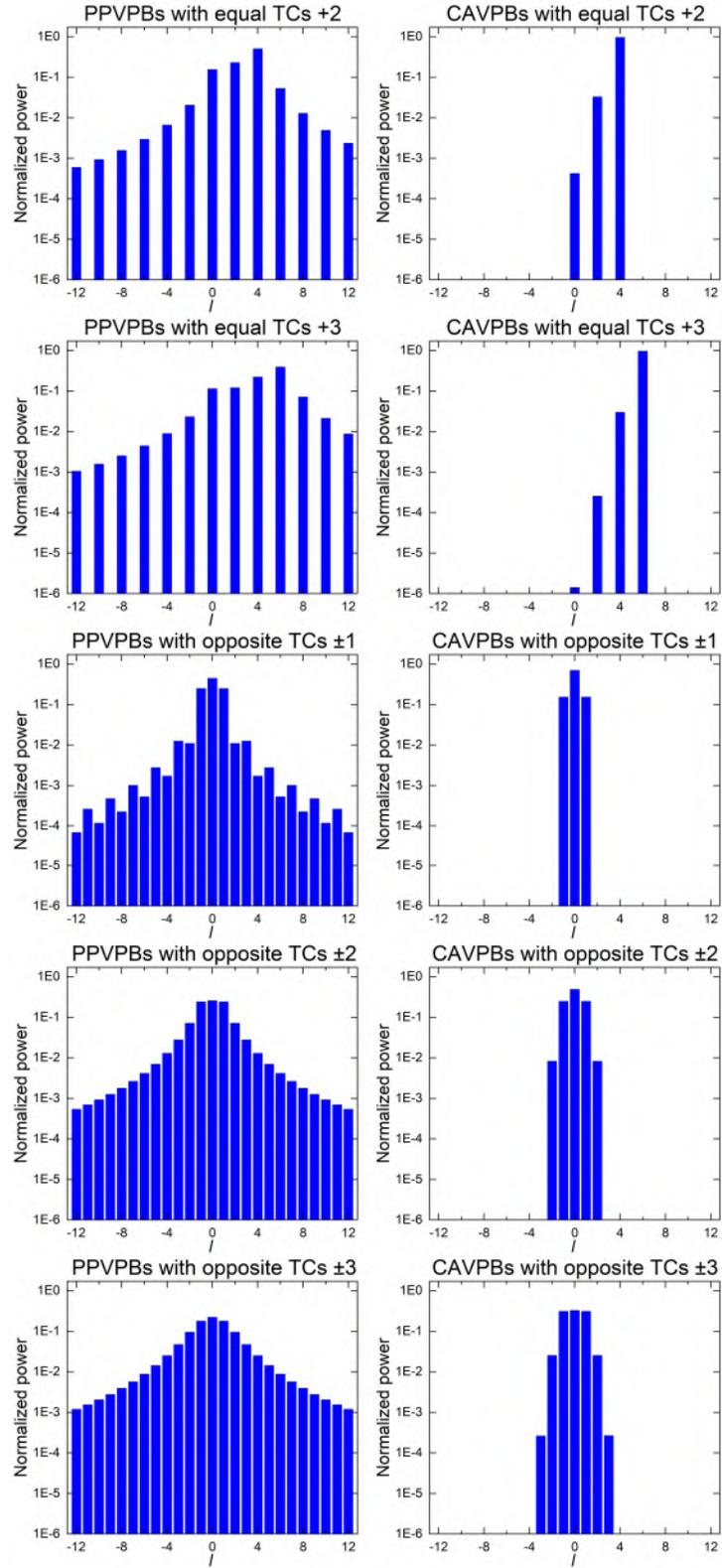

**Fig. S3. OAM spectra of the PPVPBs and CAVPBs with different TCs at the initial plane.** The parameters are $\tilde{u}_0 = 0.4$ and $w_0 = 1.63$ mm.

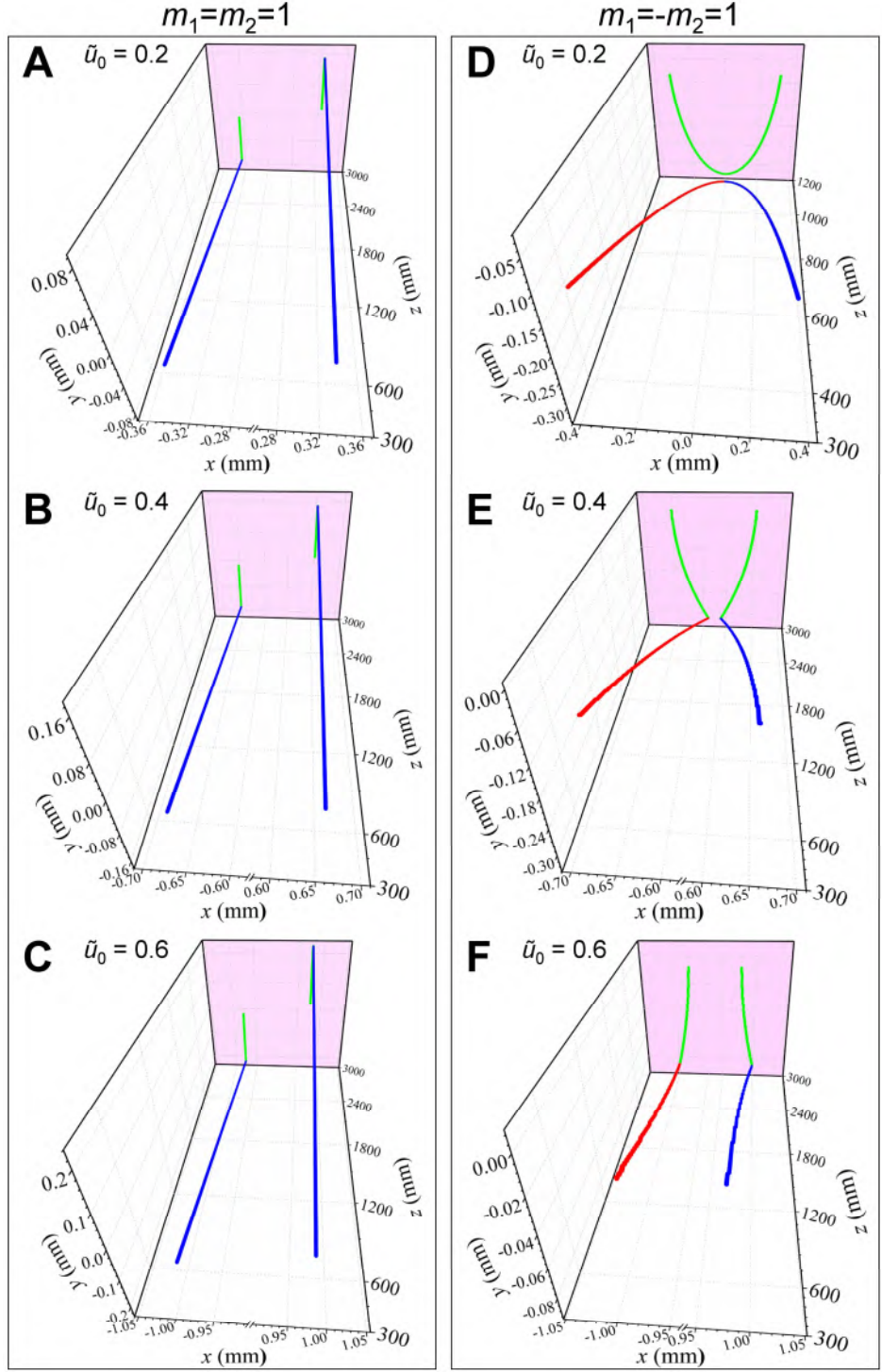

**Fig. S4. The evolution of vortex trajectories in the fields of the CAVPBs with equal and opposite TCs under different $\tilde{u}_0$ upon free space propagation.** (**A**, **B** and **C**) $m_1 = m_2 = 1$ and (**D**, **E** and **F**) $m_1 = -m_2 = 1$, and the values of $\tilde{u}_0$ are denoted in each subfigure. The blue and red lines denote, respectively, the trajectories of positive and negative vortices, and their projections in the $x$-$y$ plane are presented by the green lines. The parameter is $w_0 = 1.63$ mm.

### B. The difference between PPVPBs and CAVPBs about intensity and vortex evolutions in free space

In Fig. S4, it demonstrates the evolution of two vortices in the fields of CAVPBs. As in most of previous studies (*25*, *32*, *35*, *36*, *38*), the dynamics of the vortex pair in the fields of CAVPBs are simple for both the cases of $m_1 = m_2 = 1$ and $m_1 = -m_2 = 1$, compared to the dynamics of PPVPBs in Fig. 2. For CAVPBs, in the case of $m_1 = m_2 = 1$, the intervortex distance between two vortices increases as propagation, demonstrating the repulsion process. While in the situation of $m_1 = -m_2 = 1$, the intervortex distance between two vortices can decrease and it shows the attraction process when the initial distance is smaller than a critical value, and contrary when the initial distance is larger than the critical value the intervortex distance can also increase so that it shows the repulsion process. Meanwhile, we also plot Fig. S5 to demonstrate the similar evolutions of vortex pairs in the cases of $m_1 = m_2 = 2$ and $m_1 = -m_2 = 2$ for the fields of CAVPBs. However, in the fields of CAVPBs, from both Fig.S4 and Fig. S5, we can see that there are no fast helical and intertwined behaviors exhibited by the interaction between vortices. While, in Figs. 2 and 3 in our manuscript, in the fields of PPVPBs, such fast helical and intertwined behaviors resemble an interesting dance of vortex pairs and are very common. Such helical and intertwined behaviors induce the oscillation of the intervortex distance of the vortex pair in the dynamics of PPVPBs.

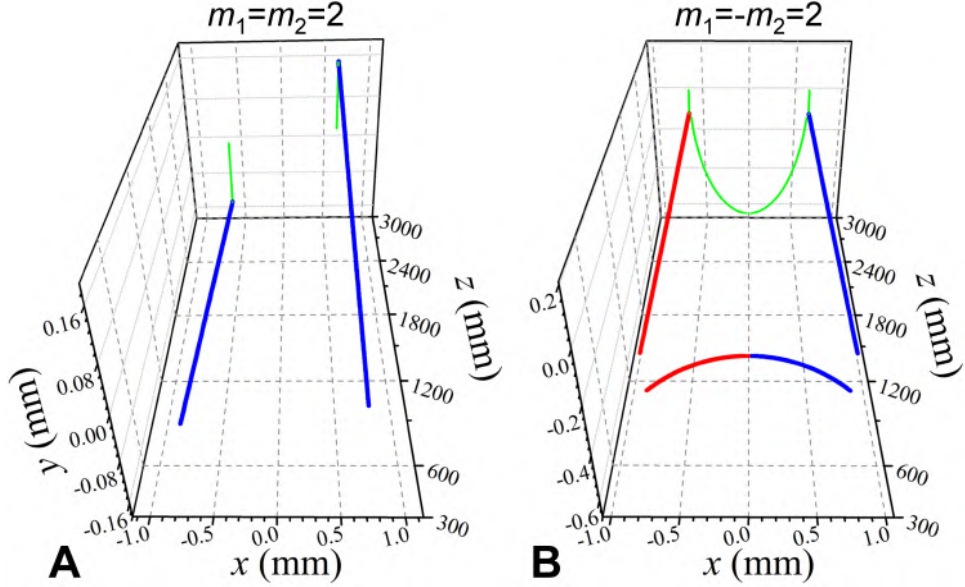

**Fig. S5. The evolution of vortex trajectories for the high-order CAVPBs in free space.** (A) $m_1 = m_2 = 2$, and (B) $m_1 = -m_2 = 2$. The blue and red lines denote, respectively, the evolution of positive and negative vortices, and their projections in the x-y plane are presented by the green lines. Other parameters are $\tilde{u}_0 = 0.4$ and $w_0 = 1.63$ mm.

In Movie S1, we present typical intensity evolutions of the PPVPBs and CAVPBs with unit TCs at different propagation distances z in free space. For the same TC +1, the intensity distributions of the PPVPBs and CAVPBs are centrosymmetric, and the intensity patterns gradually rotate anticlockwise as the value of z increases, see Movie S1(A and B). However, there are some differences between the intensity evolutions of the PPVPBs and those of the CAVPBs. For the PPVPBs with the same TC +1, as z increases, the central light intensity of the beam gradually becomes strong and an intensity peak appears, with a long rod intensity structure in the central area, see Movie S1A. For the CAVPBs with the same TC +1, with the increase of z, the intensity distributions of the beam remain unchanged, besides the anticlockwise rotation of the intensity pattern, as shown in Movie S1B. A vortex possesses a dark core at the center, where the intensity is zero (*4*, *5*). Therefore, the dynamical behaviors of vortices can be inferred from the evolution of dark regions in the light field. For the same TC +1, the two positive vortices in the PPVPBs and CAVPBs all rotate anticlockwise as a whole, as demonstrated in Movie S1(A and B). Interestingly, in the light field of the PPVPBs with the same TC +1, the positive vortex itself rotates clockwise, except for the anticlockwise rotation behavior of the vortex pair, as clearly shown in Movie S1A. For opposite TCs ±1, the intensity patterns of the PPVPBs and CAVPBs are horizontally symmetric, with a vertical line of symmetry, see Movie S1(C and D). Similarly, the positions for that pair of positive and negative vortices in the PPVPBs and CAVPBs with opposite TCs ±1 are horizontally symmetric, and this symmetry property is independent of z, as displayed in Movie S1(C and D). From Movie S1(C and D), it can be observed that the vortices in the PPVPBs behave in a very different way from those in the CAVPBs. For the PPVPBs with ±1 TCs, the positive vortex itself rotates clockwise, and the negative vortex itself rotates anticlockwise. While for the CAVPBs with ±1 TCs, the rotation behaviors of vortices cannot be seen, and vortices of ±1 TCs in this case only approach each other.

Movie S2 demonstrates typical intensity evolutions of the PPVPBs and CAVPBs with equal TCs +2 and opposite TCs ±2 under different propagation distances z in free space. Similar to the PPVPBs and CAVPBs with unit TCs, the centrosymmetric and anticlockwise rotation properties or the horizontally symmetric property of the intensity patterns can be also found in the PPVPBs and CAVPBs with equal TCs +2 or opposite TCs ±2. In Movie S2(A and B), the split of high-order vortices, the rotation of vortices themselves, and the birth and annihilation of vortices, which cannot be observed in the CAVPBs with equal TCs +2, can be clearly seen in the PPVPBs with equal TCs +2. In Movie S2D, the split of high-order vortices, and the phenomenon of opposite TCs vortices attracting each other can be seen in the CAVPBs with opposite TCs ±2. Note that the rotation of vortices themselves, which cannot be found in the CAVPBs with opposite TCs, can be observed in the PPVPBs with opposite TCs, as shown in Movie S2C. These results imply that the vortex behaviors in the PPVPBs are quite different from those in the CAVPBs.

**Movie S1.** **Typical intensity evolutions of the PPVPBs and CAVPBs under different propagation distances $z$ in free space.** (**A**) The PPVPBs with equal TCs +1, (**B**) the CAVPBs with equal TCs +1, (**C**) the PPVPBs with opposite TCs ±1, and (**D**) the CAVPBs with opposite TCs ±1. The left and middle columns are in normal and logarithmic scales, respectively. The regions marked by the black-dashed square boxes in the middle column are magnified in the corresponding figures of the right column. All the intensities are normalized. The parameters are $\tilde{u}_0 = 0.4$ and $w_0 = 1.63$ mm. The white scale bar in the left-top subfigure denotes 1.5 mm. (see https://github.com/wangligangZJU/video/blob/main/MovieS1.gif )

**Movie S2.** **Typical intensity evolutions of the high-order PPVPBs and CAVPBs at different propagation distances $z$ in free space.** (**A**) The PPVPBs with equal TCs +2, (**B**) the CAVPBs with equal TCs +2, (**C**) the PPVPBs with opposite TCs ±2, and (**D**) the CAVPBs with opposite TCs ±2. The left and middle columns are in normal and logarithmic scales, respectively. The regions marked by the black-dashed square boxes in the middle column are magnified in the corresponding figures of the right column. All the intensities are normalized. The other parameters are same as those in **Movie S1**. The white scale bar represents 1.5 mm. (see https://github.com/wangligangZJU/video/blob/main/MovieS2.gif )

## C. Hydrodynamics explanation for the motions of vortices in PPVPBs and CAVPBs upon free space propagation

Here we explain the different dynamics of vortices in PPVPBs and CAVPBs with the optical hydrodynamic picture. Using the Madelung transform on optical fields, the paraxial Helmholtz equation becomes into two coupled equations (*75*, *76*). Following the Madelung transformation by assuming the electric field as $E(x,y,z) = A(x,y,z)\exp[i\varphi(x,y,z)]$ with amplitude $A(x,y,z) \geq 0$ and phase $\varphi(x,y,z)$ and substituting it into the scalar paraxial wave equation $\nabla_\perp^2 E + i2k\frac{\partial E}{\partial z} = 0$ in free space, one obtains $\frac{\partial \varphi}{\partial z} + \frac{1}{2k}(\vec{v}\cdot\vec{v}) = \frac{\nabla_\perp^2 A}{2kA}$ and $k\frac{\partial A^2}{\partial z} + \left(\nabla_\perp \cdot (A^2\vec{v})\right) = 0$, where $\nabla_\perp^2$ and $\nabla_\perp$ are the symbols of the transverse Laplace and gradient operators, $\vec{v} = \nabla_\perp \varphi$ is the transverse velocity field of the whole beam (it is analogous to the velocity field of a fluid), and $I = A^2$ is the intensity of an optical field (like the "density" of a fluid). These two equations can be seen as the optical Bernoulli and the optical continuity equations. One can prove that the transverse velocity field in the Madelung picture is similar to the transverse Poynting vector for the linear polarization of light (*76*). From the hydrodynamic point of view, the dynamics of optical vortices is driven by two important terms: the phase gradient (i.e., the velocity field) and the intensity gradient (or the amplitude gradient). Meanwhile, one can also define the "quantum" potential $Q$ and the "quantum" force $\vec{F}$ for optical fields as follows, $Q \propto -\frac{\nabla_\perp^2 A}{2kA}$ and $\vec{F} = -\nabla_\perp Q$. One has used these concepts to successfully explain optical properties of Airy beams (*77*).

According to the recent studies (*62-64*), researchers have employed the two-dimensional hydrodynamics model to physically explain the motion of vortices, the splitting of higher-charge optical vortices, and the dynamics of vortices in complex-amplitude modulation vortex pairs. Here, we employ the same argument to explain the vortex dynamics, especially in PPVPBs, which have never considered before. According to the hydrodynamic model of an optical fluid (*62, 64*), the initial total light field, $E^i(u,v,z=0)$, consisting of multiple vortices embedded in a host Gaussian beam, can be separated into a product of two fields: $E^i(u,v,z=0) = E^i_{bg}(u,v) E^i_t(u,v)$, where $E^i_t(u,v)$ is the initial tested field of one vortex under consideration, and $E^i_{bg}(u,v)$ is the initial background field for the rest field that comprises the fields of other vortices and the host beam. In general, the total light field $E(x,y,z)$ at any propagation distance $z$ cannot be written as a product form, since the propagation of the background field at any propagation distance will couple to the propagation of the test field at $z$. Following the procedure of Ref. (*62*), here we use the evolution of the initial background field in a paraxial system approximating as the background field $E_{bg}(x,y,z)$ at any propagation distance $z$. As stated in Ref. (*62*), this approximation is reasonable at early propagation stages where the vortices are expected to be mostly circular. Thus, one can also write $E_{bg}(x,y,z) = A_{bg}(x,y,z)\exp[i\varphi_{bg}(x,y,z)]$ with $A_{bg}(x,y,z)$ the background amplitude and $\varphi_{bg}(x,y,z)$ the background phase, and the evolution of the background field obeys the paraxial diffraction equation (*66*). For example, in the cases of PPVPBs with $m_1=1$ and $m_2=-1$, the initial background field is given by $E^i_{bg}(u,v) = \exp\left(-\frac{u^2+v^2}{w_0^2}\right)\left[\frac{u+u_0+iv}{\sqrt{(u+u_0)^2+v^2}}\right]^{-1}$, and in the cases of PPVPBs with $m_1=2$ and $m_2=-2$, the initial field of the background field is consisting of one -2 TC vortex on the left side and one +1 TC vortex on the right side so that the expression for the initial background field is now given by $E^i_{bg}(u,v) = \exp\left(-\frac{u^2+v^2}{w_0^2}\right)\left[\frac{u+u_0+iv}{\sqrt{(u+u_0)^2+v^2}}\right]^{-2}\left[\frac{u-u_0+iv}{\sqrt{(u-u_0)^2+v^2}}\right]$. From the paraxial diffraction equation, one can obtain the evolution distribution of the background field $E_{bg}(x,y,z)$ at the propagation distance $z$. The transverse velocity field from the background field acting on a positive unit-charge vortex in the right side can be calculated by (*62*) $\vec{v} = k\nabla_\perp \varphi_{bg} - \vec{k}\times\nabla_\perp \ln A_{bg}$, where $k$ is the wavenumber of the laser field and $\vec{k}$ is the wavevector. Here $\vec{k}$ is simply taken as $\vec{k} \approx k\vec{z}$ and this approximation is valid only when the transverse wavevector can be neglected.

In Movie S3, we show the numerical motions of the right-side tested vortices in the background velocity fields. At short propagation distances (roughly $z<1000$ mm), in both the cases of PPVPBs with (A) $m_1=m_2=1$ and (B) $m_1=-m_2=1$, the tested vortex on the right side "surfs" in the diffraction waves from the left-side vortex, and the velocity fields (*i.e,* the circulation flow) in the background optical fluid drive quickly the local rotation and oscillation of the tested vortex during propagation and the vortex trajectories are well matched with the transverse velocity fields. At longer propagation distances, there appear some deviation between the vortex trajectories and velocity fields since we have not included the vortex tilt effect as described in Ref. (*63*), which involves the more complex calculations. However, since the oscillation and helical phenomena of vortices in PPVPBs mainly happen at short propagation distances, thus the velocity field by $\vec{v} = k\nabla_\perp \varphi_{bg} - \vec{k}\times\nabla_\perp \ln A_{bg}$ at the vortex locations is enough to explain the current vortex behavior. After the diffraction wave from the left-side vortex dissipates or becomes stable, the vortex motion will back to simply repel or attract each other. Thus, such helical and oscillation behaviors increase the repulsion effect or delay the annihilation process. As a comparison, in Movie S3(C and D) for the corresponding CAVPBs, the background velocity fields are very simple, so that there are no oscillation and helical effects in vortex motions at short propagation distances and their trajectories at far-field regions are well explained recently in Ref. (*63*) by including the vortex tilt effect.

In Movie S4, we further show the numerical motions of the right-side tested vortices in the background velocity fields. There are two +1 TC vortices coincidently located at the same initial position, these two vortices form a higher TC vortex with a +2 charge. When they split to multiple unit-charge vortices, we should plot out the locations of all these vortices since their initial conditions are the same as identical "particles". From Movie S4, in the cases of PPVPBs with (A) $m_1=m_2=2$ and (B) $m_1=-m_2=2$, there are possibly four vortices co-existing on the right-side region due to the vortex nucleation and vortex annihilation phenomena. As demonstrated in Movie S4, at short propagation distances, all the trajectories for positive vortices are considerably well matched with the transverse velocity field. The presence of the circulation flow from the right-side positive vortex further alters the local background velocity field near the right-side vortex, compared with Movie S3(A and B). Under the diffraction ripples of the background field during propagation, the tested vortices experience not only the helical and oscillating motions but also the vortex nucleation and vortex annihilation phenomena. Thus, such helical and oscillation motions in PPVPBs with opposite TCs further delay or slow down the merger process, which is reflected in smaller critical values of occurring the vortex-pair annihilation at the far-field region (or the focal plane) in Table 1. In the cases of CAVPBs, the velocity field of the background field has no ripples, so that no helical and complex motions happen there. In the case of the CAVPB with $m_1=m_2=2$, the motion of the tested vortex is coincident with the motion of the background vortex on the right side. While in the case of the CAVPB with $m_1=-m_2=2$, one of the vortices on the right side moves very slowly, and another one moves quickly along the direction of the velocity field and it will annihilate with the opposite vortex from the left side. These results for the CAVPBs agree with those in Ref. (*64*). Note that in our calculations, there is also deviation in the directions of the velocity field during the vortex nucleation or vortex annihilation, where the vortex tilt should be included (*63*) (the current calculation is already well explained the observed phenomena), and also the velocity field in Movie S4 is not suitable for generated negative TC vortices because the velocity field is given by $\vec{\upsilon} = k\nabla_\perp \varphi_{bg} + \vec{k} \times \nabla_\perp \ln A_{bg}$ for the negative-charge tested vortices.

In turn, one can also take the initial background field as a combination of the right-side positive unit-charge vortex and host Gaussian beam to investigate the motion of the left-side negative unit-charge vortex, or take the background field consisting of the right-side +2 TC vortex and one left-side -1 TC vortex embedded in the host beam for searching the dynamics of the left-side negative vortices. Therefore, the interesting vortex dancing and oscillation in PPVPBs can be well explained in the optical hydrodynamical picture and the "interactions" among vortices can be presented through the circulation flow of each background field other than the tested vortex.

**Movie S3.** **Motions of right-side tested vortices in the background velocity fields of PPVPBs and CAVPBs under different propagation distances *z* in free space.** (**A**) PPVPBs with $m_1 = m_2 = 1$, (**B**) PPVPBs with $m_1 = -m_2 = 1$, (**C**) CAVPBs with $m_1 = m_2 = 1$, and (**D**) CAVPBs with $m_1 = -m_2 = 1$. The green dots denote the locations of positive vortices, and the red arrows denote the velocity fields of the background fields. Brightness is the light intensity of background fields. All the intensities are normalized. The parameters are $\tilde{u}_0 = 0.4$ and $w_0 = 1.63$ mm. (See https://github.com/wangligangZJU/video/blob/main/MovieS3.gif )

**Movie S4.** **Motions of right-side tested vortices in the background velocity fields of high-order PPVPBs and CAVPBs under different propagation distances *z* in free space.** (**A**) PPVPBs with $m_1 = m_2 = 2$, (**B**) PPVPBs with $m_1 = -m_2 = 2$, (**C**) CAVPBs with $m_1 = m_2 = 2$, and (**D**) CAVPBs with $m_1 = -m_2 = 2$. The green and yellow dots denote, respectively, the locations of positive and negative vortices. The red arrows denote the velocity fields of the background fields. Note that the velocity field in the empty area is not shown for better displaying the velocity field in other area since it is divergent near the right-side vortex contained in the background field. Brightness is the light intensity of background fields. All the intensities are normalized. The other parameters are same as those in **Movie S3**. (See https://github.com/wangligangZJU/video/blob/main/MovieS4.gif )

## D. The evolution of intervortex distance with different values of $\tilde{u}_0$ for the PPVPBs and CAVPBs with unit vortices in free space

Fig. S6(A and B) theoretically shows the full picture on the evolution of the intervortex distance $d$ between vortices in PPVPBs under different $\tilde{u}_0$ in free space. In Fig. S6(A and B), when $m_1 = \pm m_2 = 1$, in the near-field region, the amplitude and intervortex distance of the vortex oscillation increase with the increase of $\tilde{u}_0$, and the amplitude of the vortex dance also increases as the propagation distance $z$ increases. In the far-field region, there are no vortex oscillation phenomena and the relative intervortex distances between the vortices tend to be stable, comparing with the expansion of the beam width $w(z)$. For $m_1 = m_2 = 1$, there is a fluctuation in the intervortex distances in the near field, and then the vortex spacing gradually increases with the increase of the propagation distance, and the vortex spacing gradually stabilizes in the far field, as shown in Fig. S6A. The similar dynamical behaviors can be also found in vortices of opposite TCs, as demonstrated in Fig. S6B. Interestingly, for the fields of PPVPBs with opposite TCs $m_1 = -m_2 = 1$, the intervortex spacing can not only oscillate but also decrease and finally becomes zero within one Rayleigh length under these small values of $\tilde{u}_0$ (such as $\tilde{u}_0=0.2$ and $\tilde{u}_0=0.3$), see the black and red lines in Fig. S6B, indicating the vortex helical, intertwined behaviors and the vortex annihilation of positive and negative vortices in PPVPBs with opposite TCs. While in the cases of $\tilde{u}_0 = 0.4$, $\tilde{u}_0 = 0.5$ and $\tilde{u}_0 = 0.6$, the positive and negative vortices survive in the far-field region and always remain the non-zero vortex spacing, see the blue, green and purple lines in Fig. S6B. These results mean that there is a critical value of $\tilde{u}_0$ between 0.3 and 0.4, at which the opposite TCs vortices can annihilate each other in the far field.

For the detail comparison, both Fig. S6C and Fig. S6D present the intervortex distances for vortices in CAVPBs upon free space propagation. For the CAVPBs with $m_1 = m_2 = 1$, the relative distance $d/w(z)$ between the two vortices maintains invariant under different values of $\tilde{u}_0$ as seen in Fig. S6C, although the absolute value of $d$ increases as the host beam width $w(z)$ increases. When $m_1 = -m_2 = 1$, the two vortices in the CAVPBs may approach each other, leading to the gradually decreasing vortex spacing as shown in Fig. S6D. In this case, if the value of $\tilde{u}_0$ is small enough (such as $\tilde{u}_0=0.2, 0.3$ and $0.4$), the vortex spacing can be reduced to zero over a finite distance, indicating the attraction process of two vortices. From Fig. S6D, one can also find that when $\tilde{u}_0 = 0.5$, the intervortex distances between two vortices will tend to be zero as $z$ goes to the infinity of free space. As the value of $\tilde{u}_0$ is larger than 0.5, the absolute value of $d$ increases since $w(z)$ increases, and this also indicates the repulsion process of two vortices. However, comparing Fig. S6(A and B) with Fig. S6(C and D), one can see an essential difference that the value of $d/w(z)$ has no oscillation effect for the cases of CAVPBs, and the oscillating behaviors in the near-field regions tells us the rich interaction between two vortices in PPVPBs that is absent in CAVPBs. Through a careful comparison between Fig. S6A and Fig. S6C, one can find that in the case of $m_1 = m_2 = 1$ and under the same $\tilde{u}_0$, the vortex spacing in the far fields of PPVPBs is always greater than that of CAVPBs, which indicates the stronger repulsion process in the fields of PPVPBs. Comparing Fig. S6B and Fig. S6D, we also observe that for $m_1 = -m_2 = 1$ and under the same $\tilde{u}_0$, the opposite vortices in PPVPBs can survive (i.e. keeping the non-zero vortex spacing) over longer distances than the corresponding cases in CAVPBs. Obviously, the oscillation and intertwined behaviors between two vortices significantly influence the process of vortex-pair annihilation and prolong the survival range of opposite vortices in PPVPBs.

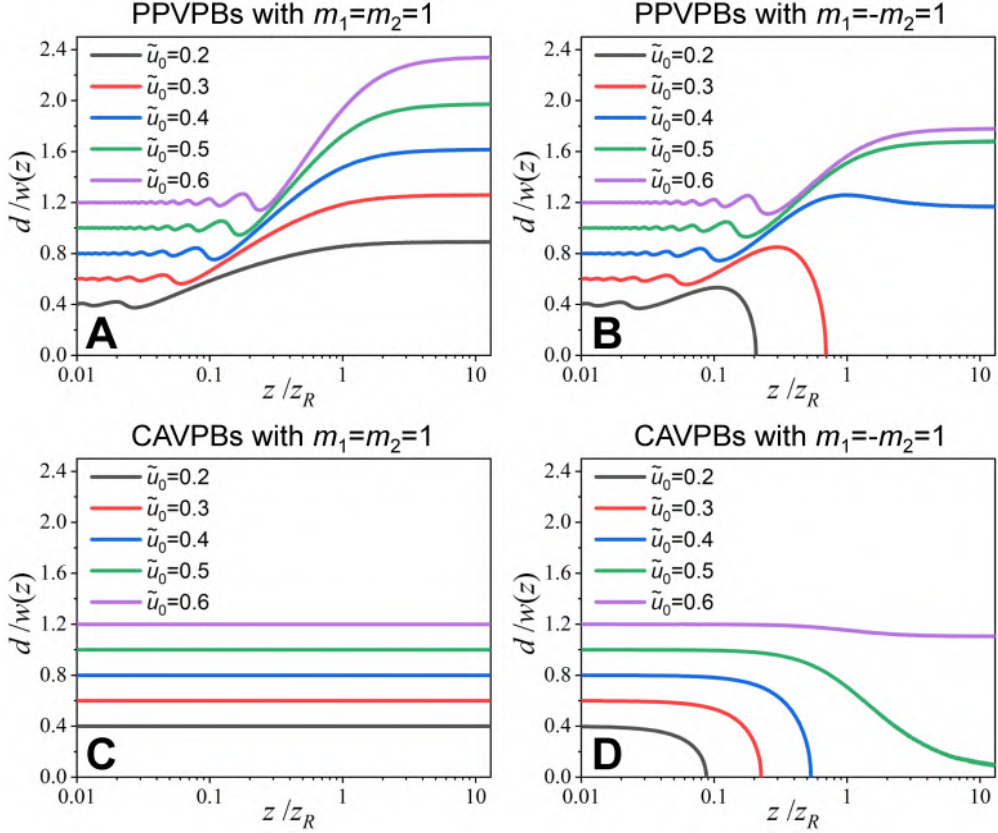

**Fig. S6. The intervortex distance $d$ between the two vortices as a function of the propagation distance $z$ under different relative off-axis distances $\tilde{u}_0$ in free space.** (**A** and **B**) The PPVPBs with $m_1 = m_2 = 1$ and $m_1 = -m_2 = 1$, respectively; (**C** and **D**) the CAVPBs with $m_1 = m_2 = 1$ and $m_1 = -m_2 = 1$, respectively. Note that the propagation distance $z$ is normalized by the Rayleigh length $z_R$ of the host Gaussian beam and the value of the intervortex distance $d$ is also rescaled by $w(z)$ the beam width of the host beam at $z$.

### E. Additional theoretical results of the vortex trajectories for the PPVPBs with $m_1 = m_2 = 2$ in free space

Figure S7 shows the trajectories of vortices for the PPVPBs with equal TCs $m_1 = m_2 = 2$ in the case of free space. From Fig. S7, it can be found that there are entanglement behaviors between four separate positive vortices, and these vortices do helicoidal motions with the increase of the propagation distance $z$. Interestingly, there are the generation and annihilation processes for pairs of positive and negative vortices in the fields of the PPVPBs upon free space propagation. As the value of $\tilde{u}_0$ increases, the vortex entanglement phenomena, the vortex helical behaviors, and the generation and annihilation processes of vortices become obvious. Thus, the parameter $\tilde{u}_0$ can be used to control the dynamics of the PPVPBs.

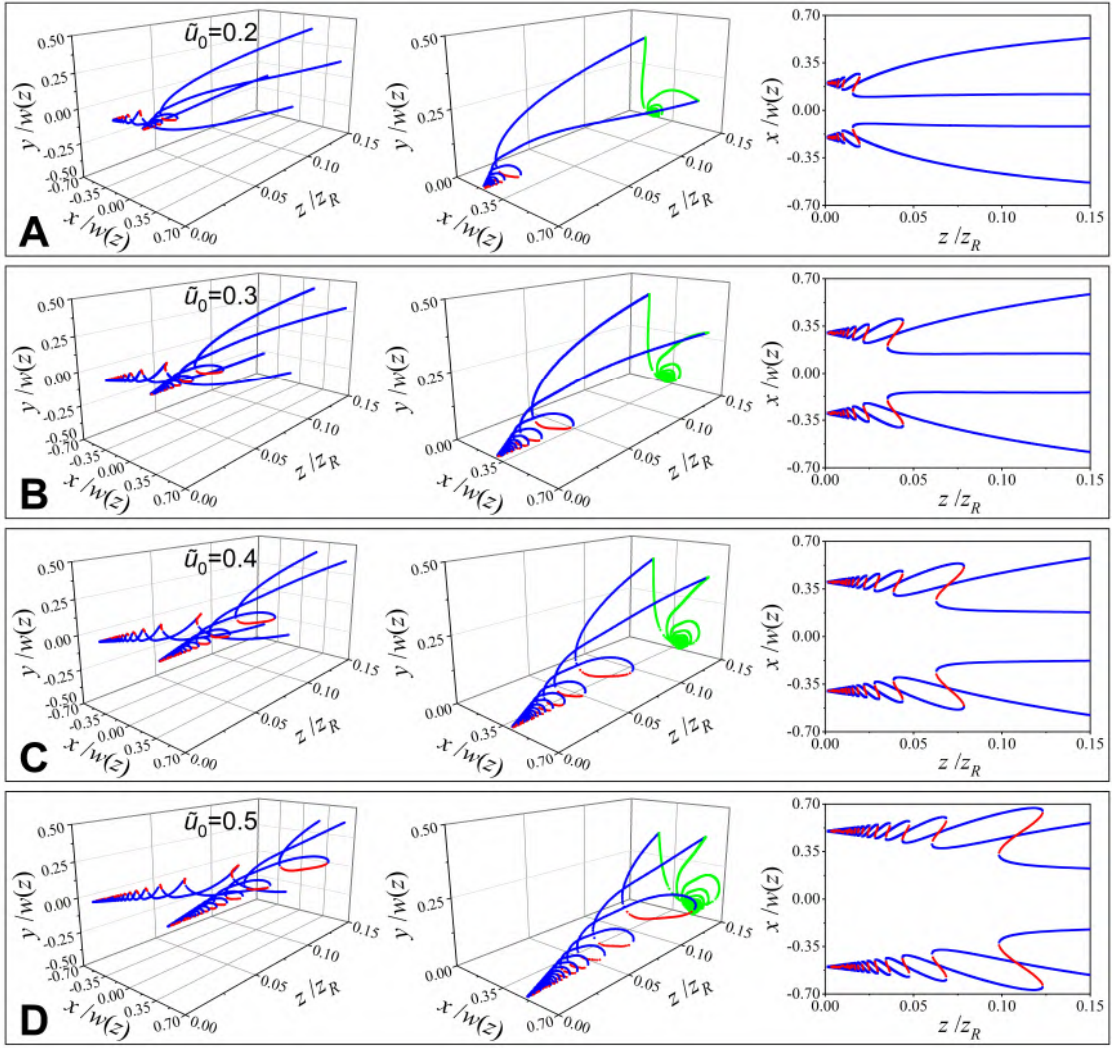

**Fig. S7. Theoretical evolution of phase singularities in the fields of the PPVPBs with the same TC $m_1 = m_2 = 2$ upon free space propagation under different relative off-axis distance parameters.** (**A**) $\tilde{u}_0 = 0.2$, (**B**) $\tilde{u}_0 = 0.3$, (**C**) $\tilde{u}_0 = 0.4$, and (**D**) $\tilde{u}_0 = 0.5$. The middle and the right subfigures are the enlarging parts and the projections in the *x-z* plane of the left subfigures, respectively. The blue and red lines denote, respectively, the trajectories of positive and negative vortices, and their projections in the *x-y* plane are presented by the green lines.

### F. Theoretical evolution of vortices for the PPVPBs with $m_1 = -m_2 = 2$ upon free space propagation

Figure S8 demonstrates the vortex trajectories for the PPVPBs with opposite TCs $m_1 = -m_2 = 2$ in free space. It is observed that there are two pairs of vortices with ±1 TCs, and their helical behaviors and entanglement phenomena become significant with the increase of the relative off-axis distance $\tilde{u}_0$. Therefore, the relative off-axis distance parameter can be used as a control parameter for controlling the dynamical behaviors of the PPVPBs. In addition, when the parameter $\tilde{u}_0$ is small enough, such as $\tilde{u}_0 = 0.2$, there are annihilation phenomena between one pair of positive and negative unit vortices related to the initial light field in near-field regions, as displayed in Fig. S8A. It can be inferred that a smaller $\tilde{u}_0$ can contribute to the annihilation behaviors of opposite TCs vortices.

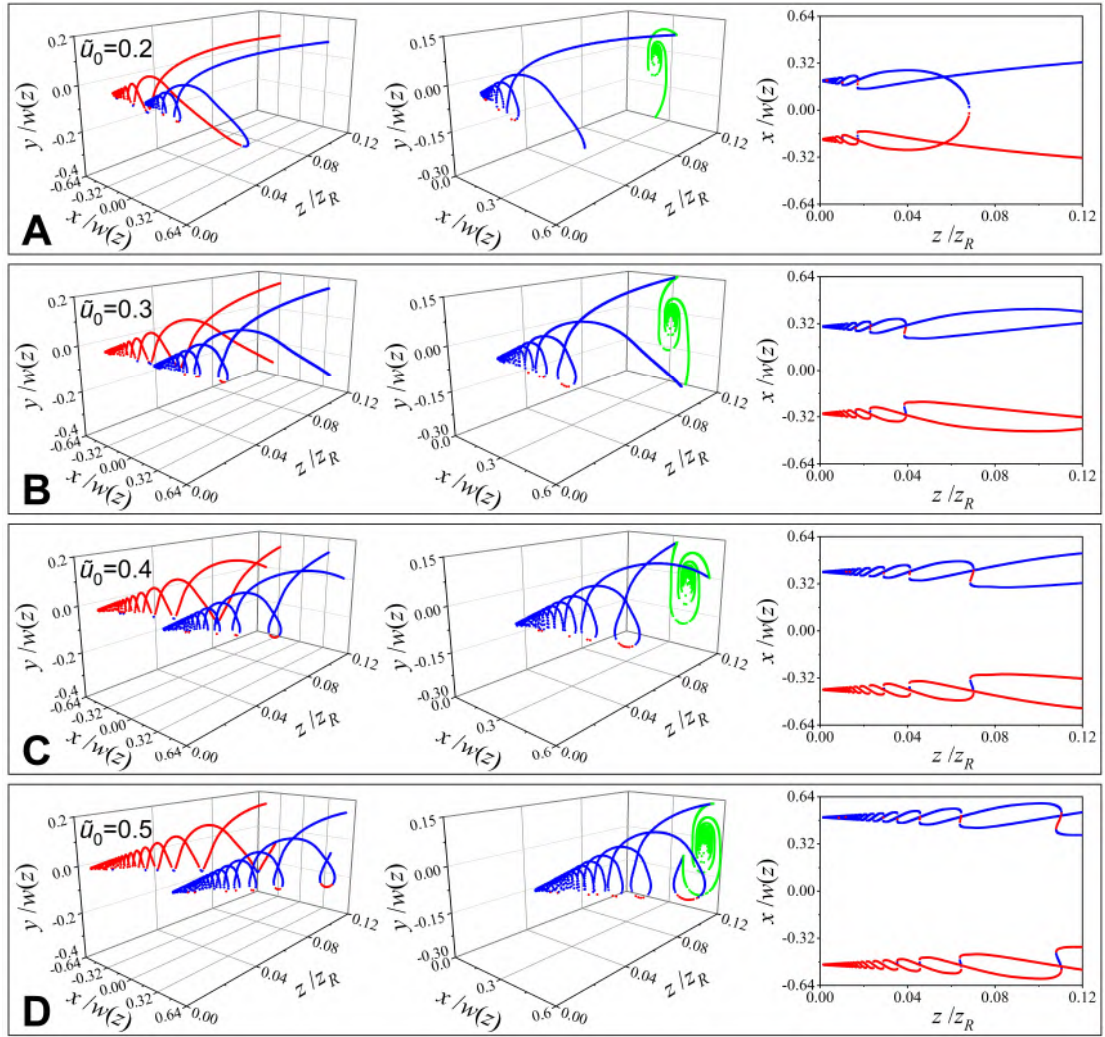

**Fig. S8. Theoretical trajectories of phase singularities for the PPVPBs with opposite TCs $m_1 = -m_2 = 2$ in free space.** The other explanations and parameters are same as those in **Fig. S7**.

### G. The vortex trajectories of the CAVPBs with different opposite TCs and relative off-axis distances in a 2-*f* lens system

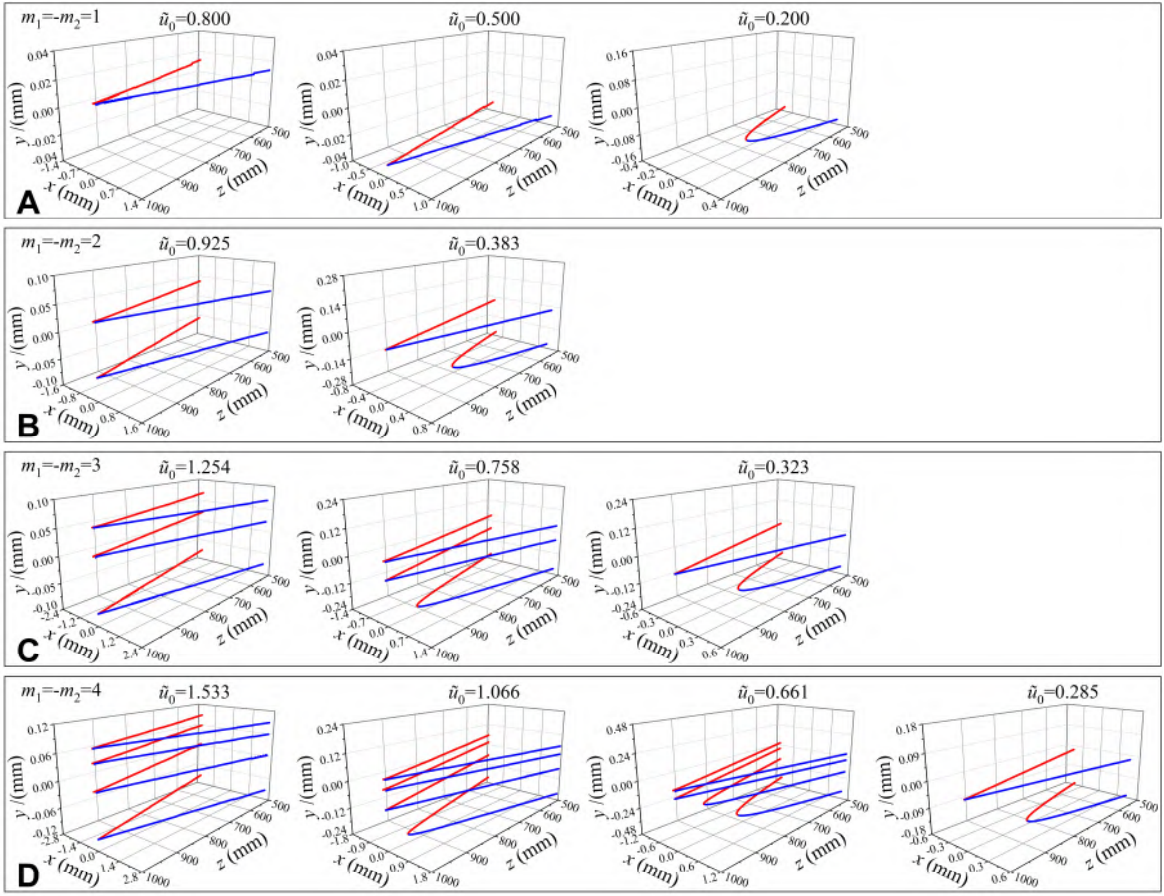

**Fig. S9**. **Evolution of vortex trajectories in different-order CAVPBs with opposite TCs in the 2-*f* focusing system.** (**A**) The influence of the relative off-axis distance $\tilde{u}_0$ on the vortex-trajectory evolution, where the opposite vortices happen to annihilate each other at the focusing plane when $\tilde{u}_0 = 0.5$. (**B**, **C**, and **D**) The vortex-trajectory evolutions at various critical values of $\tilde{u}_0$ in different-order CAVPBs, in which each plot corresponds to the situation that one pair of opposite vortices annihilate each other at the focusing plane. Here, the focal length of the 2-*f* focusing system is $f = 500$ mm and the beam parameter is also taken to be $w_0 = 1.63$ mm.

Figure S9 shows the annihilation processes of vortices for the CAVPBs with opposite TCs in a 2-*f* lens system with the focal length $f = 500$ mm. For the CAVPBs with $m_1 = -m_2 = 1$, when the relative off-axis distance ($\tilde{u}_0$) is big enough, such as $\tilde{u}_0 = 0.800$, the vortex-pair can survive to the back focal plane of the 2-*f* lens system, as displayed in Fig. S9A. As $\tilde{u}_0$ decreases, the vortex-pair tends to meet and annihilate. When $\tilde{u}_0 = 0.500$, the vortex pair undergoes annihilation upon reaching the back focal plane. If $\tilde{u}_0 = 0.200$, the vortex-pair will annihilate prior to reaching the back focal plane. These results indicate that $\tilde{u}_0 = 0.500$ is the critical value of the vortex-pair annihilation in CAVPBs for $m_1 = -m_2 = 1$: when $\tilde{u}_0 < 0.500$, the annihilation behavior of the vortex-pair can be seen; while $\tilde{u}_0 > 0.500$, the vortex-pair will survive all the way to the back focal plane or the far field. Due to the unstable properties of high-order vortices in propagation, for $m_1 = -m_2 = 2, 3$ and 4, the pair of vortices with ±2, ±3 and ±4 TCs at the initial light field for the CAVPBs will split into two, three and four pairs of opposite unit vortices upon propagation, respectively. Therefore, as demonstrated in Fig. S9(B, C and D), there are two, three and four critical values about $\tilde{u}_0$ of the vortex-pair annihilation for $m_1 = -m_2 = 2, 3$ and 4, respectively. From Fig. S9, it is worth noting that the vortex-pair dancing behaviors, which can be found in the PPVPBs as clearly demonstrated in Fig. 4, have never been observed in the CAVPBs.

**H. The change of the intervortex distance for the fields of PPVPBs with $m_1 = -m_2 = 1$ in free space under different relative off-axis distances $\tilde{u}_0$, which correspond to the different situations in Fig. 4A.**

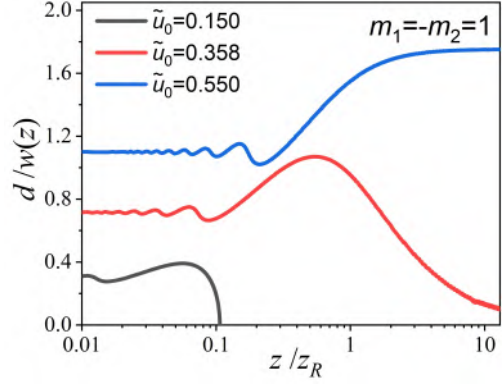

**Fig. S10. The intervortex distance *d* between the two vortices as a function of the propagation distance *z* under different relative off-axis distances $\tilde{u}_0$ in free space for the fields of PPVPBs with $m_1 = -m_2 = 1$.** Note that the propagation distance *z* is rescaled by the Rayleigh length $z_R$ of the host Gaussian beam and the value of *d* is rescaled by *w(z)*, which is the beam width of the host beam at *z*.

Here we illustrate the similarity of light propagations in the 2-*f* lens system and the far-field region of free space. For the PPVPBs with $m_1 = -m_2 = 1$, when $\tilde{u}_0 = 0.55$, the positive and negative vortices undergo oscillations and survive at the back focal plane, see the leftmost subfigure in Fig. 4A. This feature is consistent with the non-zero intervortex spacing in free space, see the blue line in Fig. S10. When $\tilde{u}_0 = 0.358$, the positive and negative vortices coincidentally merge and annihilate each other at the back focal plane, see the middle subgraph of Fig. 4A, and this predicts that the intervortex spacing *d* tends to be zero at the infinity of free space as shown by the red line in Fig. S10, or it indicates that the vortex pair will merge together at the infinity of free space. When $\tilde{u}_0 = 0.150$, in this situation the vortex-pair can merge prior to reaching the focal plane, as seen in the rightmost subgraph of Fig. 4A, and this shows that the two vortices with opposite TCs merge and annihilate each other at a certain propagation distance in free space and the separation distance between the vortices goes to be zero at a certain distance as demonstrated by the black line in Fig. S10. Thus, the dynamics of vortices in the 2-*f* focusing system can effectively and conveniently demonstrate the key characteristics of vortex interactions among vortex pairs within the limited propagation distance.

## I. The impact of different widths of the host Gaussian beam on the evolutions of a single off-axial pure-phase vortex, a single off-axial complex-amplitude vortex, PPVPBs, and CAVPBs

To demonstrate the amplitude modulation of the host Gaussian beam on PPVPBs and CAVPBs, we can first investigate the Gaussian modulation of the host beam on a single pure-phase vortex and a single complex-amplitude vortex for the sake of simplicity. In this situation, we can set $m_2$ to zero and take different values of $m_1$ in Eqs. (S5) and (S6). Figures S11-S14 show the typical intensity and phase evolutions and the distance from the origin for a single vortex in free space. For the case of a single unit vortex embedded in a host Gaussian beam, as the propagation distance increases, the vortex "rotates" around the origin of the transverse plane. Interestingly, a single positive vortex "rotates" counterclockwise, while a single negative vortex "rotates" clockwise, with a rotation angle of $\arctan(z/z_R)$ and a distance of $\tilde{u}_0 w(z)$ from the origin, as shown in Figs. S11 and S12. Here, $w(z) = w_0\sqrt{1+(z/z_R)^2}$ is the host beam width at the propagation distance *z*. When we read out the coordinate data of these vortex locations, we find that these vortex locations in fact move along the ±*y* direction (i.e., the straight line of $x = 0.4w_0$ in these cases (*25*, *78*)). There is the same law for a single high-order vortex in both pure-phase and complex-complex cases, see Fig. S13. This can be explained since both a single pure-phase vortex and a single complex-amplitude vortex are affected only by the same background field (i.e., the same host Gaussian beam). From the perspective of vortex dynamics, the behaviors of a single pure-phase vortex and a single complex-amplitude vortex are similar, and there is no oscillation in the distance between the origin and the vortex position, see Fig. S12. In the case of a single high-order vortex, such as +2 or -2 TC, the rotation properties of the intensity and phase patterns, and the distance between the origin and the vortex location of the single pure-phase vortex are also similar to those of the single complex-amplitude vortex, as illustrated in Figs. S13 and S14, which are consistent with the single unit vortex. These results indicate that the host Gaussian beam plays the same role in controlling the dynamics of a single embedded vortex for both pure-phase and complex-amplitude cases.

However, for the evolution of light intensity in Figs. S11 and S13, they are different for the pure-phase and complex-amplitude cases. Considering the expansion of the host beam width $w(z)$, it can be observed that the light intensity profile of a single pure-phase vortex not only rotates but also diffracts/spreads out during propagation, while for a single complex-amplitude vortex it rotates rigidly and does not diffract/spread out, see Figs. S11 and S13. The larger the TC values, the more evident the diffraction effect in the pure phase cases. Such diffraction patterns in the pure-phase vortex situation as the additional background fields will influence the vortex behaviors of other vortices in the pure-phase vortex pairs.

To better reveal the amplitude modulation of the host Gaussian beam on the vortex dynamics, we can introduce an expansion parameter $b$ of the Gaussian beam waist to adjust the initial light field of VPBs. In this situation, the light fields of PPVPBs and CAVPBs at the initial plane now can be given by

$$E_{\text{PPVPB}}^{b}(u,v) = F(u,v)\exp\left[-\frac{u^2+v^2}{(bw_0)^2}\right], \tag{S9}$$

and

$$E_{\text{CAVPB}}^{b}(u,v) = M(u,v)\exp\left[-\frac{u^2+v^2}{(bw_0)^2}\right], \tag{S10}$$

where $b$ is a positive real number and represents the amplification factor of the waist $w_0$ for the host Gaussian beam. The larger the value of $b$, the more the Gaussian beam approximates a plane wave (i.e, its amplitude of the host beam is more approximately normalized for both kinds of VPBs). Figures S15-S18 present the vortex trajectories and intervortex distance for PPVPBs and CAVPBs under different $b$ in free space. In Fig. S15, for PPVPBs with unit TCs, the vortex spacing slightly decreases as the value of $b$ increases. A large $b$ can contribute to the annihilation of opposite vortices, see Fig. S15F. From Fig. S15(B, C, E and F), it is found that the oscillation effect of the intervortex distance at short distances is nearly overlapped for different values of $b$. Clearly, the behaviour of vortex dynamics at short distances is dominated mainly due to the interaction of two vortices. The large difference at longer distances is due to the contribution of the host beam. When the value of $\tilde{u}_0$ is small (such as $\tilde{u}_0 = 0.2$ and $0.4$), the evolution trajectory of vortex spacing under $b = 10$ almost completely overlaps with that under $b = 5$. There are also similar phenomena observed in Fig. S16 for CAVPBs with unit TCs. Based on these results, the host Gaussian beam with $b = 10$ can be regarded as a plane wave in the situations of small $\tilde{u}_0$, and the amplitude of the host beam is approximately normalized. Interestingly, in Figs. S15-S18, with the increase of $b$, vortex oscillation and helical behaviours, which cannot be observed in the CAVPBs, can be clearly seen in the PPVPBs. From Figs. S15 and S17, we can see that the host Gaussian beam has less affected on the oscillation dynamics of pure-phase vortex pairs as the value of $b$ increases. While in the cases of CAVPBs with equal TCs, the smaller $b$ can drive the faster motion of vortices (see Fig. S16(A, B and C) and Fig. S18(A and B)); in contrast, in the cases of CAVPBs with opposite TCs, the smaller $b$ slows down the annihilation process. In fact, the dynamics of vortex pair in the complex-amplitude cases has been physically explained from the optical hydrodynamics (*63*). According to the above discussion and Ref. (*63*), we can see that the host Gaussian beam play a limited role in controlling the dynamic behaviors of embedded vortices.

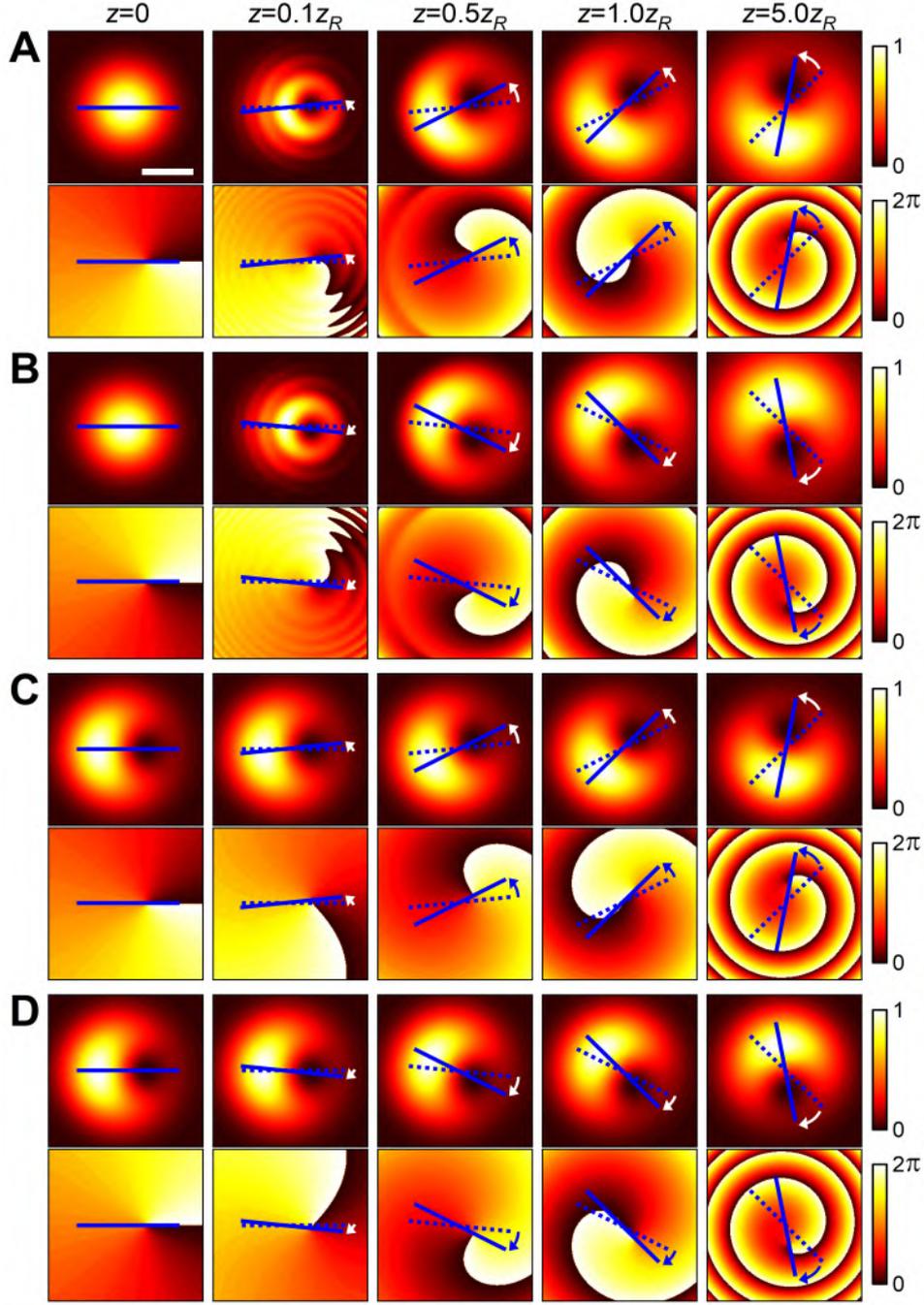

**Fig. S11. Typical intensity (upper) and phase (lower) evolutions of a single unit vortex for both PPVPBs and CAVPBs under different propagation distances $z$ in free space.** (**A**) PPVPBs with $m_1 = 1$, $m_2 = 0$, (**B**) PPVPBs with $m_1 = -1$, $m_2 = 0$, (**C**) CAVPBs with $m_1 = 1$, $m_2 = 0$, and (**D**) CAVPBs with $m_1 = -1$, $m_2 = 0$. Here, the blue solid and dashed lines to indicate the azimuthal positions of the vortices at the current and previous positions, respectively, for better showing the rotation phenomenon of intensity and phase patterns at different $z$. The notation $z$ on the top of each column indicates the same propagation distance, where $z_R$ denotes the Rayleigh length of the host Gaussian beam. Other parameters are $\tilde{u}_0 = 0.4$ and $w_0 = 1.63$ mm. The white scale bar represents 1 $w(z)$, where $w(z)$ is the beam width of the host beam at $z$. In the situations of $z = 0$, $0.1z_R$, $0.5z_R$, $1.0z_R$ and $5.0z_R$, the vortex locations $(x, y)$ in both (**A**) and (**C**) are $(0.40w_0, 0)$, $(0.40w_0, 0.04w_0)$, $(0.40w_0, 0.20w_0)$, $(0.40w_0, 0.40w_0)$ and $(0.40w_0, 2.00w_0)$, respectively. In the cases of $z = 0$, $0.1z_R$, $0.5z_R$, $1.0z_R$ and $5.0z_R$, the vortex locations $(x, y)$ in both (**B**) and (**D**) are $(0.40w_0, 0)$, $(0.40w_0, -0.04w_0)$, $(0.40w_0, -0.20w_0)$, $(0.40w_0, -0.40w_0)$ and $(0.40w_0, -2.00w_0)$, respectively.

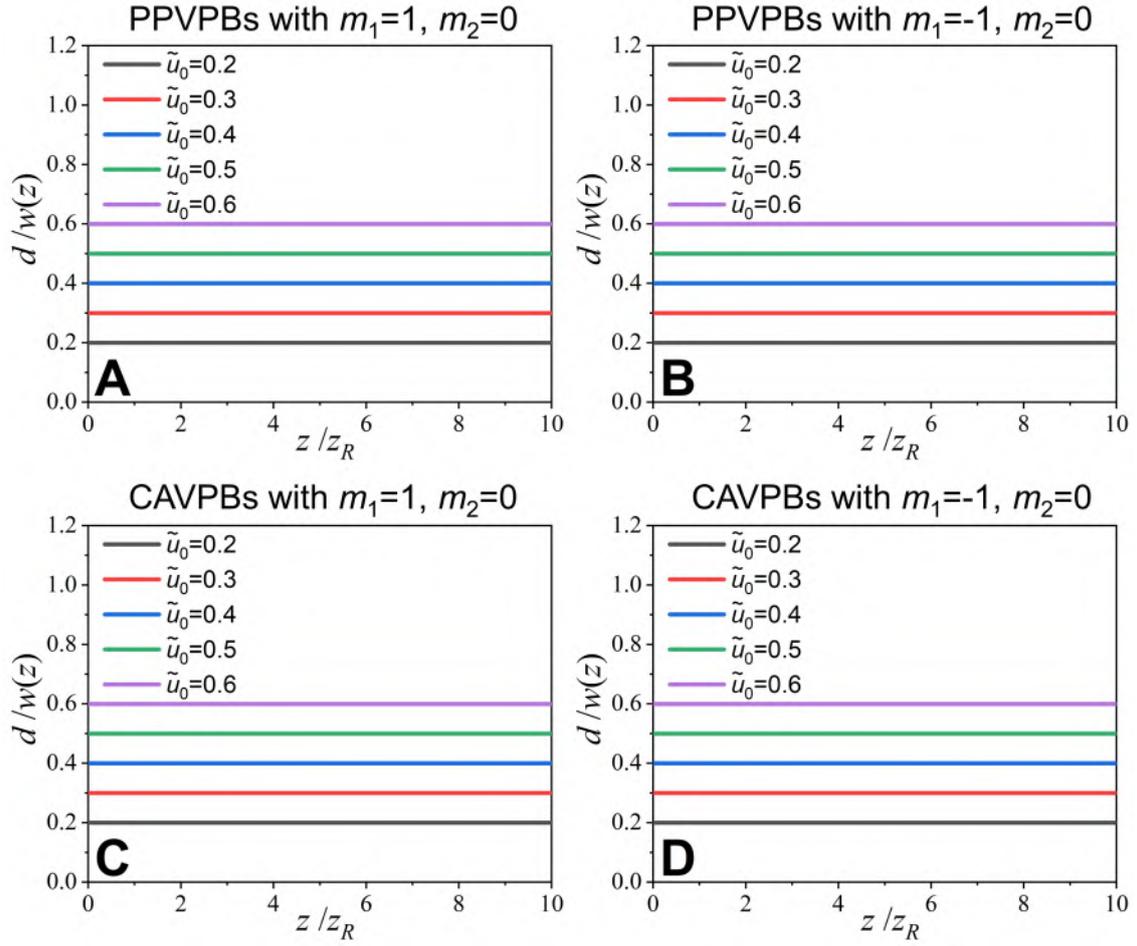

**Fig. S12. The distance $d$ between the origin and the unit vortex location as a function of the propagation distance $z$ under different relative off-axis distances $\tilde{u}_0$ in free space.** (**A** and **B**) The PPVPBs with $m_1 = 1$, $m_2 = 0$ and $m_1 = -1$, $m_2 = 0$, respectively; (**C** and **D**) the CAVPBs with $m_1 = 1$, $m_2 = 0$ and $m_1 = -1$, $m_2 = 0$, respectively. Note that the propagation distance $z$ is normalized by the Rayleigh length $z_R$ of the host Gaussian beam and the value of the vortex distance $d$ is also rescaled by $w(z)$ the beam width of the host beam at $z$.

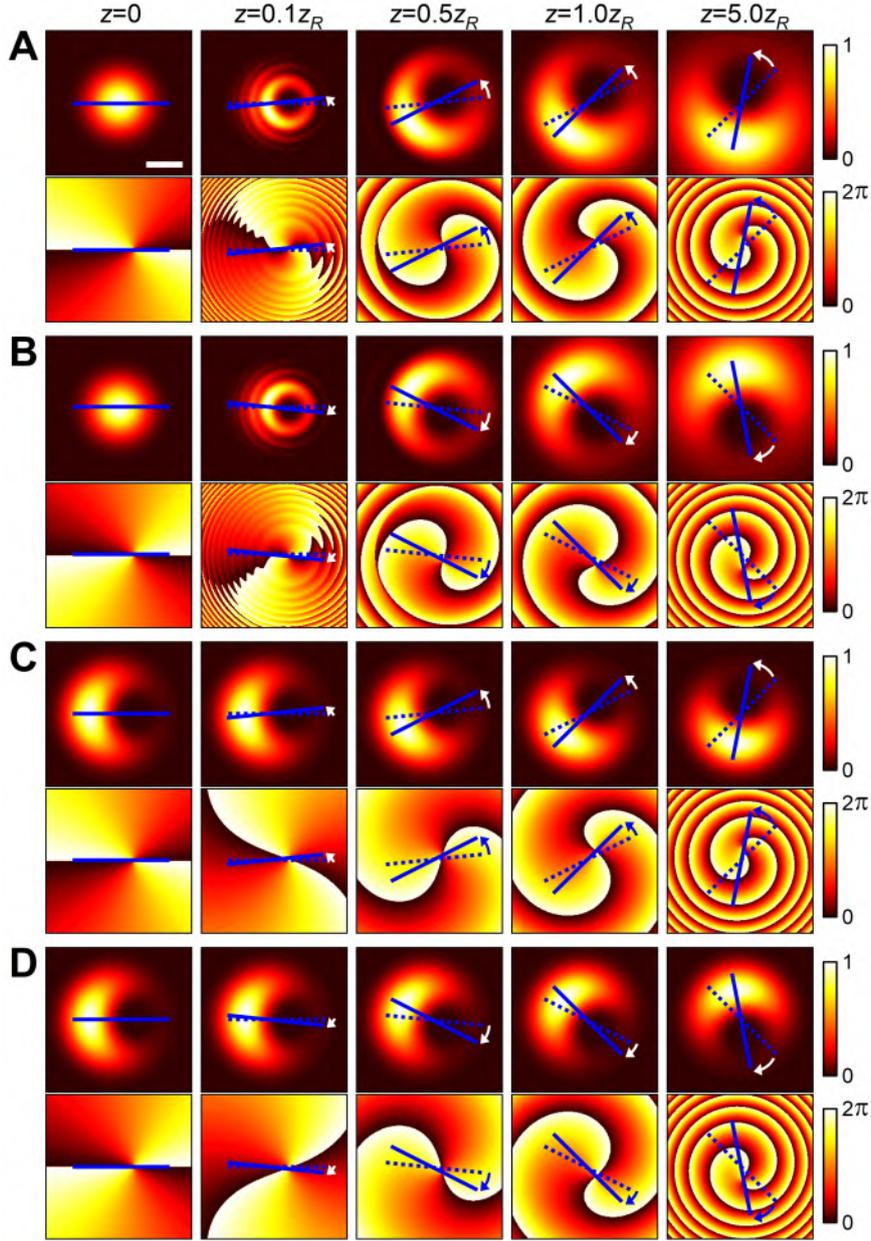

**Fig. S13. Typical intensity (upper) and phase (lower) evolutions of a single high-order vortex for both PPVPBs and CAVPBs under different propagation distances $z$ in free space.** (**A**) PPVPBs with $m_1 = 2$, $m_2 = 0$, (**B**) PPVPBs with $m_1 = -2$, $m_2 = 0$, (**C**) CAVPBs with $m_1 = 2$, $m_2 = 0$, and (**D**) CAVPBs with $m_1 = -2$, $m_2 = 0$. Here, the blue solid and dashed lines to indicate the azimuthal positions of the vortices at the current and previous positions, respectively, for better showing the rotation phenomenon of intensity and phase patterns at different $z$. The notation $z$ on the top of each column indicates the same propagation distance, where $z_R$ denotes the Rayleigh length of the host Gaussian beam. Other parameters are $\tilde{u}_0 = 0.4$ and $w_0 = 1.63$ mm. The white scale bar represents 1 $w(z)$, where $w(z)$ is the beam width of the host beam at $z$. In the situations of $z = 0$, $0.1z_R$, $0.5z_R$, $1.0z_R$ and $5.0z_R$, the vortex locations $(x, y)$ in both (**A**) and (**C**) are $(0.40w_0, 0)$, $(0.40w_0, 0.04w_0)$, $(0.40w_0, 0.20w_0)$, $(0.40w_0, 0.40w_0)$ and $(0.40w_0, 2.00w_0)$, respectively. In the cases of $z = 0$, $0.1z_R$, $0.5z_R$, $1.0z_R$ and $5.0z_R$, the vortex locations $(x, y)$ in both (**B**) and (**D**) are $(0.40w_0, 0)$, $(0.40w_0, -0.04w_0)$, $(0.40w_0, -0.20w_0)$, $(0.40w_0, -0.40w_0)$ and $(0.40w_0, -2.00w_0)$, respectively.

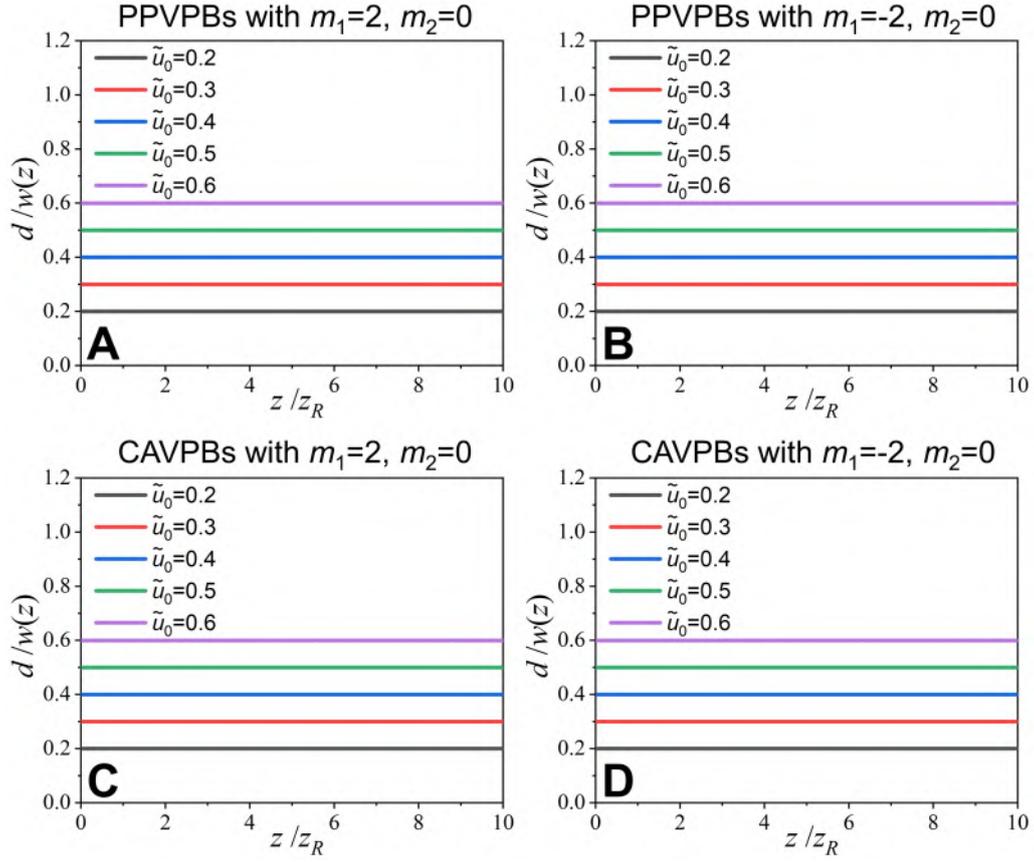

**Fig. S14. The distance *d* between the origin and the high-order vortex location as a function of the propagation distance *z* under different relative off-axis distances $\tilde{u}_0$ in free space.** (**A** and **B**) The PPVPBs with $m_1 = 2$, $m_2 = 0$ and $m_1 = -2$, $m_2 = 0$, respectively; (**C** and **D**) the CAVPBs with $m_1 = 2$, $m_2 = 0$ and $m_1 = -2$, $m_2 = 0$, respectively. Note that the propagation distance *z* is normalized by the Rayleigh length $z_R$ of the host Gaussian beam and the value of the vortex distance *d* is also rescaled by $w(z)$ the beam width of the host beam at *z*.

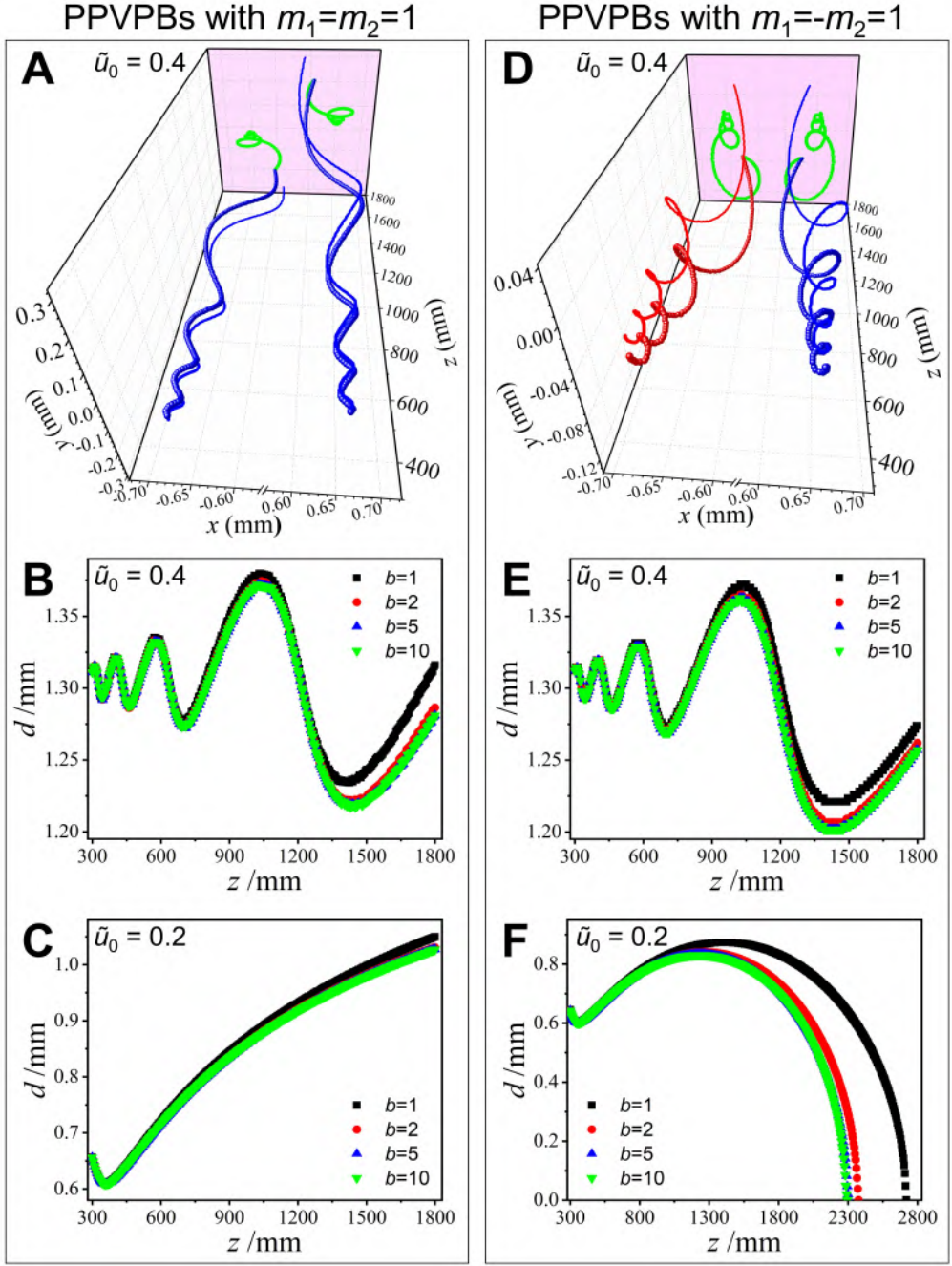

**Fig. S15. The vortex trajectories and intervortex distance for PPVPBs with unit TCs under different $\tilde{u}_0$ and $b$ upon free space propagation.** (**A** and **D**) The vortex trajectories for (**A**) $m_1 = m_2 = 1$, and (**D**) $m_1 = -m_2 = 1$ with $\tilde{u}_0 = 0.4$, $b = 10$. The blue and red dots denote, respectively, the evolution of positive and negative vortices and their projections are shown by the green dots in the $x$-$y$ planes. The corresponding thin solid lines are the cases of $b = 1$. (**B, C, E** and **F**) Evolution of the intervortex distance along the propagation distance for (**B**) $m_1 = m_2 = 1$, $\tilde{u}_0 = 0.4$, (**C**) $m_1 = m_2 = 1$, $\tilde{u}_0 = 0.2$, (**E**) $m_1 = -m_2 = 1$, $\tilde{u}_0 = 0.4$, and (**F**) $m_1 = -m_2 = 1$, $\tilde{u}_0 = 0.2$, respectively, under different $b$. The parameter is $w_0 = 1.63$ mm.

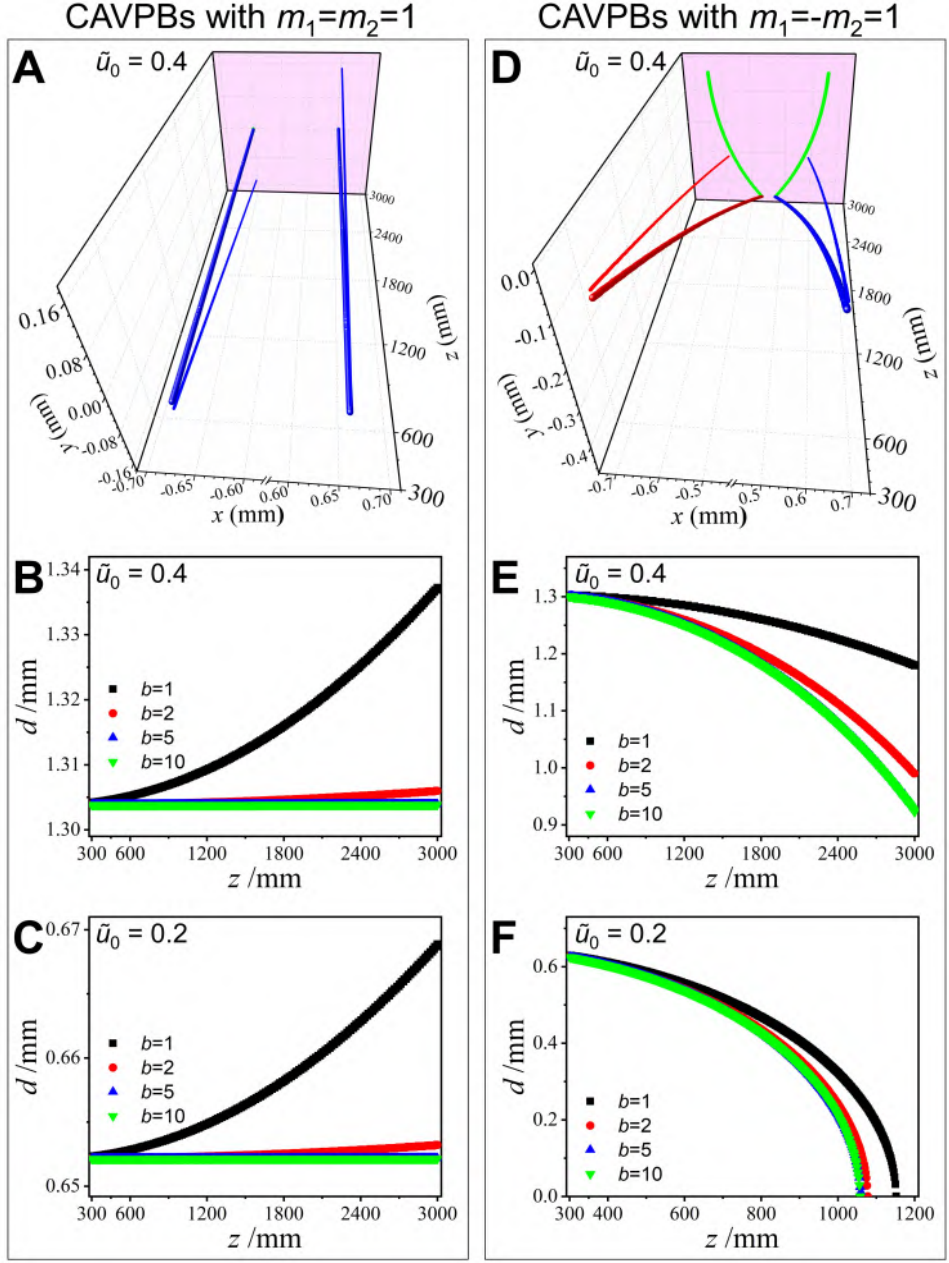

**Fig. S16. The vortex trajectories and intervortex distance for CAVPBs with unit TCs under different $\tilde{u}_0$ and $b$ upon free space propagation.** (**A** and **D**) The vortex trajectories for (**A**) $m_1 = m_2 = 1$, and (**D**) $m_1 = -m_2 = 1$ with $\tilde{u}_0 = 0.4$, $b = 10$. The blue and red dots denote, respectively, the evolution of positive and negative vortices and their projections are shown by the green dots in the $x$-$y$ planes. The corresponding thin solid lines are the cases of $b = 1$. (**B, C, E** and **F**) Evolution of the intervortex distance along the propagation distance for (**B**) $m_1 = m_2 = 1$, $\tilde{u}_0 = 0.4$, (**C**) $m_1 = m_2 = 1$, $\tilde{u}_0 = 0.2$, (**E**) $m_1 = -m_2 = 1$, $\tilde{u}_0 = 0.4$, and (**F**) $m_1 = -m_2 = 1$, $\tilde{u}_0 = 0.2$, respectively, under different $b$. The parameter is $w_0 = 1.63$ mm.

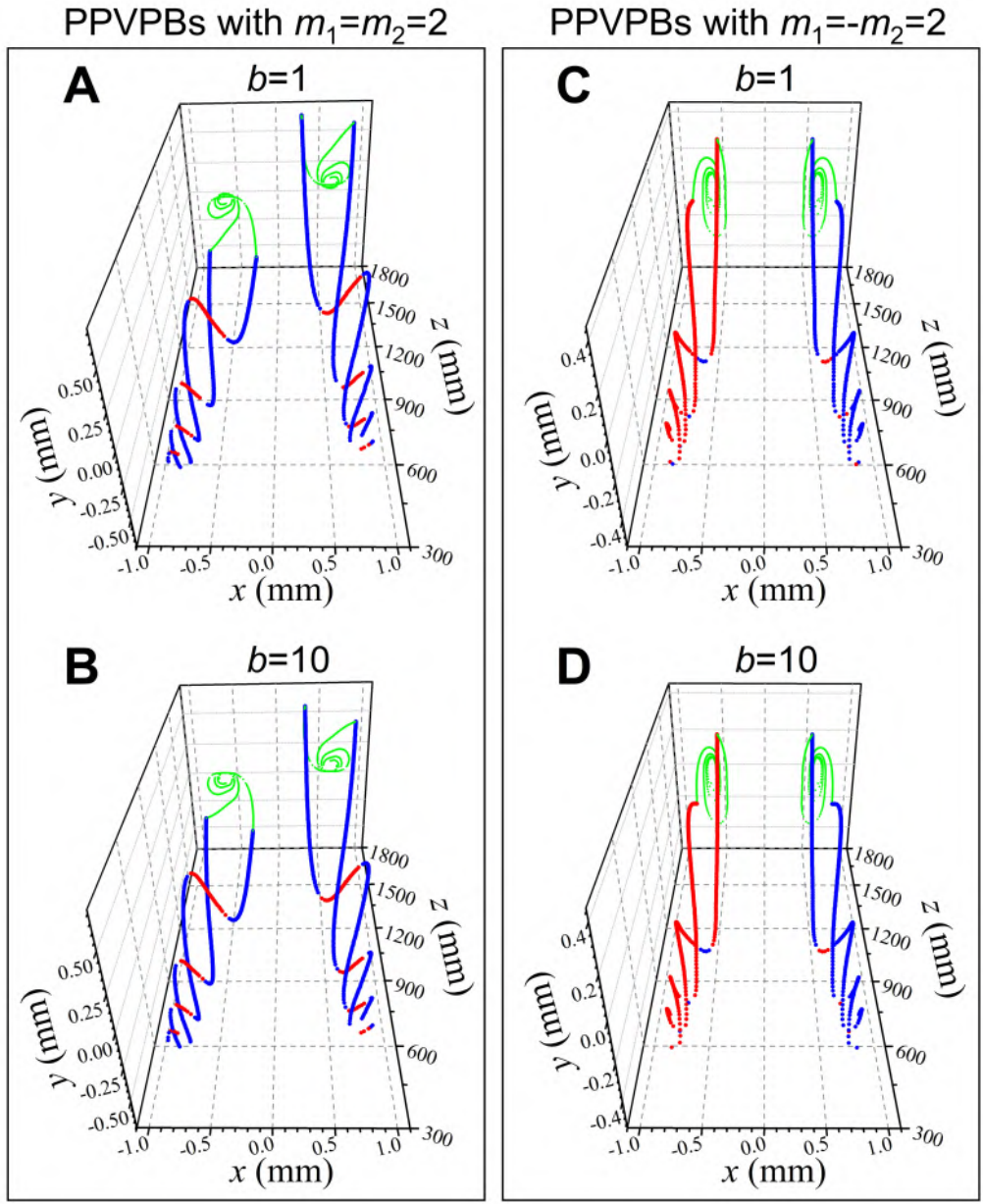

**Fig. S17. The evolution of vortex trajectories in the fields of PPVPBs with high-order TCs under different *b* upon free space propagation.** (**A** and **B**) $m_1 = m_2 = 2$ and (**C** and **D**) $m_1 = -m_2 = 2$, and the values of *b* are denoted in each subfigure. The blue and red lines denote, respectively, the trajectories of positive and negative vortices, and their projections in the *x-y* plane are presented by the green lines. Other parameters are $\tilde{u}_0 = 0.4$ and $w_0 = 1.63$ mm.

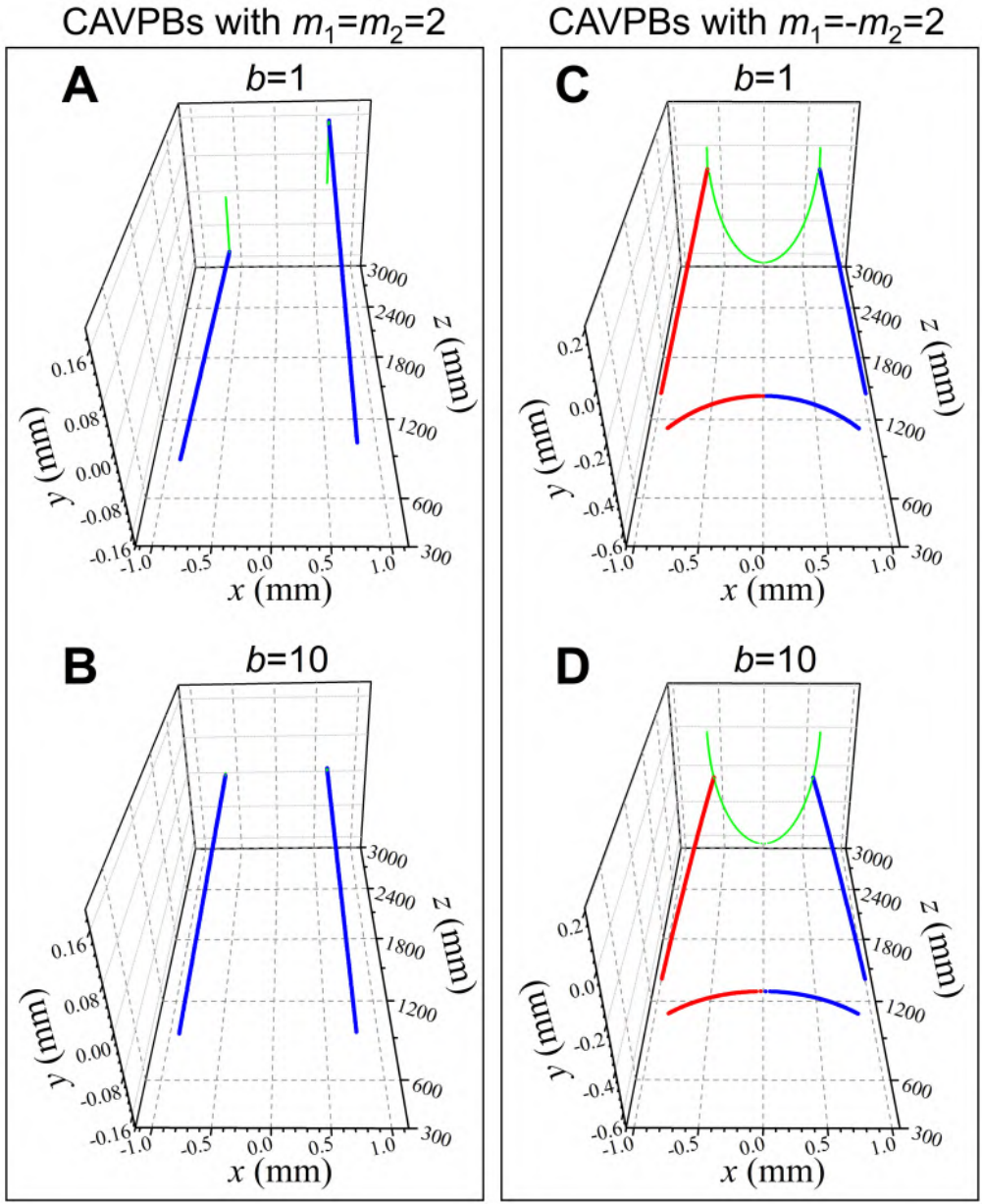

**Fig. S18. The evolution of vortex trajectories in the fields of CAVPBs with high-order TCs under different *b* upon free space propagation.** (**A** and **B**) $m_1 = m_2 = 2$ and (**C** and **D**) $m_1 = -m_2 = 2$, and the values of *b* are denoted in each subfigure. The blue and red lines denote, respectively, the trajectories of positive and negative vortices, and their projections in the *x-y* plane are presented by the green lines. Other parameters are $\tilde{u}_0 = 0.4$ and $w_0 = 1.63$ mm.